\documentclass[11pt,a4paper]{article}

\pdfoutput=1

\def\beq{\begin{equation}}
\def\eeq{\end{equation}}
\def\beqa{\begin{eqnarray}}
\def\eeqa{\end{eqnarray}}

\usepackage[pdftex]{graphicx}

\usepackage{jcappub}
\title{Rest frame of bubble nucleation}

\author[a,c]{Jaume Garriga,}
\author[b]{Sugumi Kanno,}
\author[c]{Takahiro Tanaka}
\affiliation[a]{Departament de F{\'\i}sica Fonamental i \\
Institut de Ci{\`e}ncies del Cosmos, 
Universitat de Barcelona,\\
Mart{\'\i}\ i Franqu{\`e}s 1, 08028 Barcelona, Spain}
\affiliation[b]{Institute of Cosmology, Department of Physics and Astronomy, 
Tufts University, Medford, Massachusetts 02155, USA}
\affiliation[c]{ Yukawa Institute for Theoretical Physics, Kyoto University,
Kyoto 606-8502, Japan}

\abstract{Vacuum bubbles nucleate at rest with a certain critical size and subsequently expand. But what selects {the rest frame of nucleation?} This question has been recently addressed in \cite{GKSSV} in the context of Schwinger pair production in 1+1 dimensions, by using a model detector in order to probe the nucleated pairs. {The analysis in \cite{GKSSV}  showed that, for a constant external electric field, the adiabatic ``in" vacuum of charged particles is Lorentz invariant, {and in this} case pairs tend to nucleate preferentially at rest with respect to the detector. Here, we sharpen this picture by showing that  the typical relative velocity between the frame of nucleation and that of the detector is at most of order $\Delta v \sim S_E^{-1/3} \ll 1$.  Here, $S_E \gg 1$ is the action of the instanton describing pair creation. The bound $\Delta v$ coincides with the minimum uncertainty in the velocity of a non-relativistic charged particle embedded in a constant electric field. A velocity of order $\Delta v$ is reached after a time interval of order $\Delta t \sim S_E^{-1/3} r_0\ll r_0$ past the turning point in the semiclassical trajectory,  where $r_0$ is the size of the instanton. If the interaction takes place in the vicinity of the turning point, the semiclassical description of collision does not apply. Nonetheless, we find that {even in this case} there is still a strong asymmetry in the momentum transferred from the nucleated particles to the detector, in the direction of expansion after the turning point. We conclude that the correlation between the rest frame of nucleation and that of the detector is exceedingly sharp.} }

\begin{document}
\maketitle

\section{Introduction}

False vacuum decay at zero temperature was first described in the context of field theory in a classic paper by Voloshin, Kobzarev and Okun \cite{VKO}. {These authors proposed that, as a result of quantum tunneling, true vacuum bubbles would nucleate at a certain rate per unit volume. The bubbles would nucleate at rest, with a critical size $r_0$, and then expand with constant proper acceleration $r_0^{-1}$, due to the pressure difference between the false and the true vacuum}. 

On the other hand, { it was also pointed out in \cite{VKO} that, due to the Lorentz invariance of the false vacuum,} bubbles would not have any preferred reference frame in which to nucleate at rest.
At first, this observation seemed to suggest that the total decay rate per unit volume should include an integral over the Lorentz group, in order to account for all possible rest frames {of nucleation}. 
Of course, such integral would be divergent, or at least cut-off dependent if a regulator was imposed. 
However, soon after this idea was proposed, Coleman \cite{Coleman} developed the instanton approach to 
false vacuum decay, showing that the decay rate is finite -- and that integration over the Lorentz group does not play any role in calculating it
\footnote{The reason is that instanton in this case is $O(4)$ invariant, and its analytic continuation (describing the bubble after nucleation) is invariant with respect to Lorentz boosts. Because of that, the final state in the asymptotic future is independent of the rest frame in which the critical bubble nucleates. Integrating over the Lorentz group would then amount to overcounting the final states. For a recent discussion of related issues, see \cite{Dvali,GSV,Dine:2012tj,GV12}.}.
Coleman's argument, however, {did} not shed much light on a related and somewhat mysterious aspect of bubble nucleation, which requires clarification.
If it is true that a long lived metastable false vacuum is approximately Lorentz invariant, what is it that determines the  rest frame in which the critical bubble nucleates? 

\begin{figure}
\begin{center}
\vspace{-2cm}
\includegraphics[width=14cm]{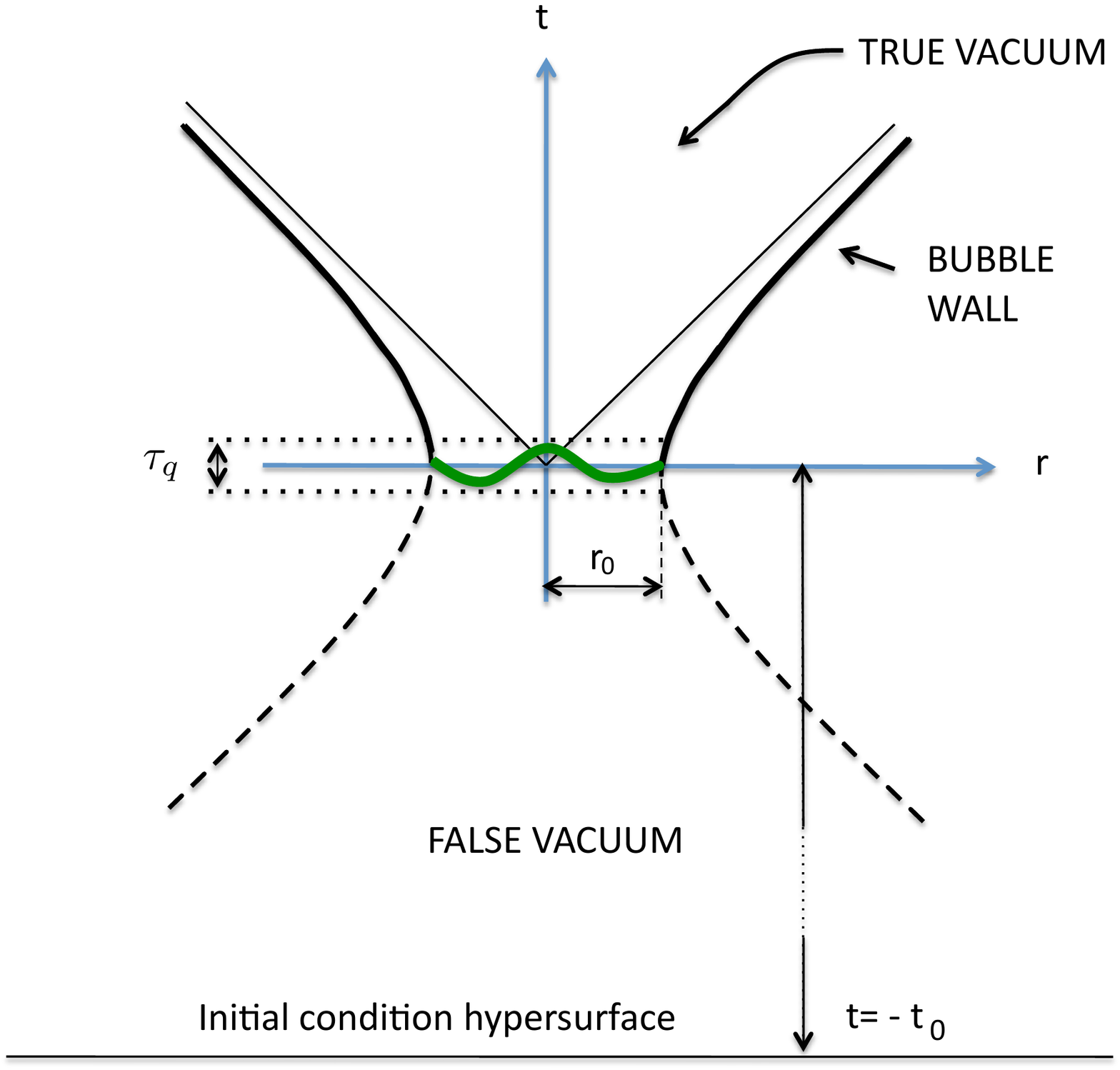}
\vspace{-1.5cm}
\caption{Diagram illustrating the nucleation of a bubble of true vacuum. The system is prepared to be in the metastable false vacuum at some early time $t=-t_0$. A bubble of size $r_0$ nucleates at rest at time $t=0$, and subsequently expands. The semiclassical picture of an expanding bubble is valid for $t\gtrsim \tau_q$.
 \label{bubble}}
\end{center}
\end{figure}

{It should be noted that the process of bubble nucleation cannot itself be Lorentz invariant. To illustrate this point, we may consider  the limit when the bubble walls are thin compared to the bubble size. 
In this case,} the trajectory 
of a vacuum bubble of radius $r$ as a function of time $t$ is given by \cite{VKO,Coleman} (see Fig. \ref{bubble})
\begin{equation}
r^2-t^2=r_0^2.    \label{trajectory}
\end{equation}
Eq.  (\ref{trajectory}) is invariant under Lorentz boosts, and describes a bubble which contracts from infinite
size to the minimum size $r_0$, and then expands again to infinity.
However, only the expanding part of this trajectory is relevant for describing vacum decay. 
Indeed, we are assuming that the system is prepared in the false vacuum in the far past, remaining in that state for a very long period of time. A tunneling event occurs at the time which we here denote as $t=0$, forming a critical bubble, which then expands according to (\ref{trajectory}).  Nucleation is a quantum process (indicated in Fig. \ref{bubble}  by a wavy line), and therefore we cannot say that it takes place exactly at the turning point hypersurface $t=0$. 
Rather, the semiclassical picture of an expanding bubble is valid only after a certain time $\tau_q$ has ellapsed, 
\begin{equation}
t \gtrsim \tau_q>0.
\end{equation}
As we shall see, in the regime where the action of the instanton describing vacuum decay is large, we have $\tau_q \ll r_0$. In this sense, it is still quite accurate to say that nucleation takes place on a $t \approx const.$ hypersurface. 

\begin{figure}
\begin{center}
\vspace{-2cm}
\includegraphics[width=14cm]{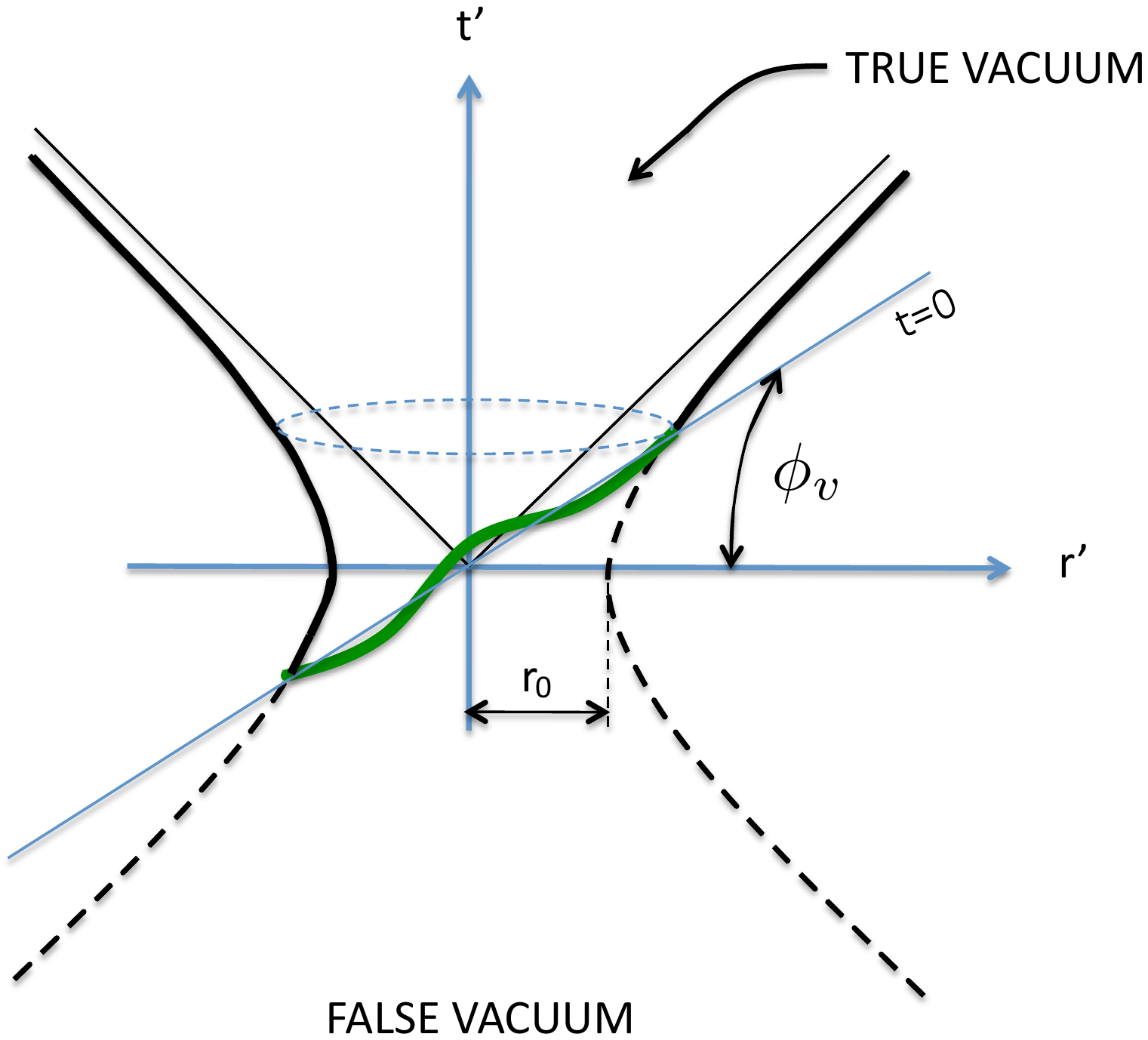}
\vspace{-1.5cm}
\caption{Bubble nucleation in a reference frame $S'$ which  moves at high velocity $v=\tanh \phi_v$ with respect to the frame of nucleation. Observers in the boosted frame will see a piece of the bubble appear at time $t'\approx -r_0\sinh\phi_v$, moving very fast opposite to the boost direction. At $t'=0$ the bubble wall has a somewhat 
awkward shape: it consists of one hemisphere attached to a fuzzy interface between false and true vacuum, which cannot be described semiclassically. 
The process of formation lasts untile the time  $t'\approx +r_0\sinh\phi_v$, when the semiclassical bubble wall closes into a full sphere.
 \label{boost}}
\end{center}
\end{figure}

Since only the expanding branch of (\ref{trajectory}) is relevant,  the actual process of vacuum decay is not at all Lorentz invariant. 
In a reference frame $S'$ which  moves at high velocity $v=\tanh \phi_v$ with respect to the rest frame of nucleation, things look rather different (see Fig. \ref{boost}). Observers in the new frame will see a piece of the bubble appear at time $t'\approx -r_0\sinh\phi_v$, moving very fast in the boost direction. The bubble then comes to a halt at $t'=0$. At that time the bubble wall presents a somewhat awkward shape: it consists of one hemisphere attached to a rather fuzzy interface between false and true vacuum, which cannot be described semiclassically. The nucleation process does not conclude until the time  $t'\approx +r_0\sinh\phi_v$, when the semiclassical bubble wall wraps the full sphere.

Given its impact on the kinematics of bubble formation, it is somewhat surprising that investigation of the frame of nucleation has been neglected for over three decades.  Recently, however, this issue was addressed in Ref. \cite{GKSSV}. The strategy which was adopted there was to consider a model detector which interacts with the bubbles, so as to reveal their state of motion at the time of interaction. Several plausibe scenarios were anticipated to possibly emerge from this study: (A) It could be that the frame of nucleation simply coincides with the rest frame of the detectors.  In other words, each detector will see bubbles forming at rest in her own rest frame.  (B) It is conceivable that the contracting part of the bubble history is not completely cut off and can be at least partially observed.  (C) A third possibility is that the frame of nucleation is influenced by how the decaying false vacuum was set up. 

The investigation in \cite{GKSSV} was done by using pair production of a charged scalar field $\phi$ in a constant electric field as a model for bubble nucleation in (1+1) dimensions. 
To take care of option  (C),  the initial hypersurface where the quantum state for the field $\phi$ is prepared was taken to the infinite past $t\to  -\infty$. In this idealized situation, it was shown that the ``in" vacuum is Lorentz invariant\footnote{This does not follow trivially from the boost invariance of the electric field, since the ``in" vacuum must be defined in a given frame and in a specific gauge.}, and therefore cannot determine a preferred frame of nucleation.

\begin{figure}
\begin{center}
\vspace{-2cm}
\includegraphics[width=14cm]{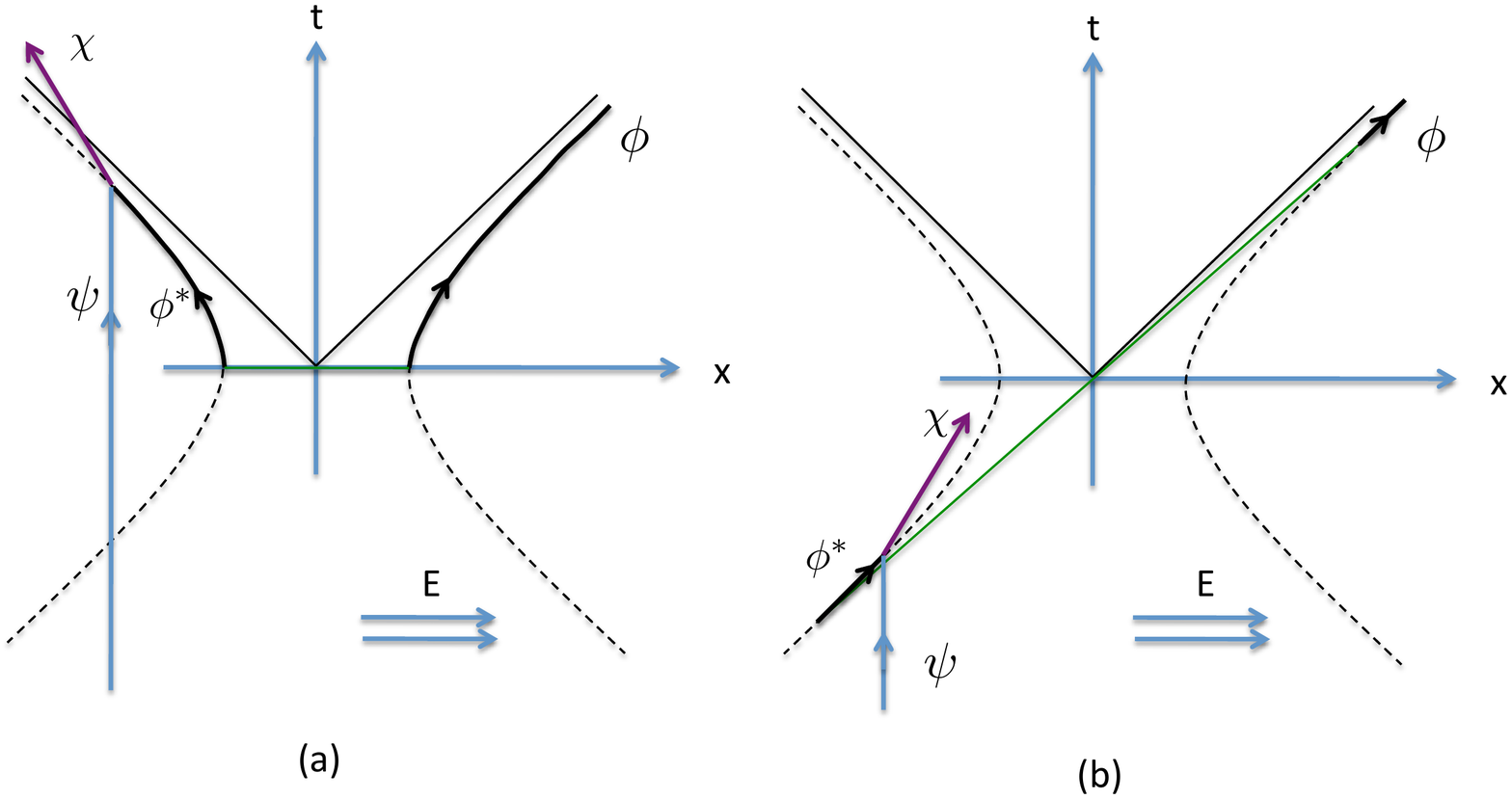}
\vspace{-1.5cm}
\caption{The ``detector" particle $\psi$ can anihilate the $\phi$ antiparticle through the vertex (\ref{vertex}), producing a neutral particle $\chi$. Kinematically, there are at most two opportunities for the interaction to take place along the hyperbolic trajectory of $\phi$.  If the pair nucleates in the rest frame of the detector particle $\psi$, as in (a), then the collision will take place in the expanding branch of the hyperbola, and the momentum of the decay product $\chi$ will be negative $p<0$. On the other hand, if the pair nucleates in a frame which is highly boosted with respect to the detector, then there is a good chance that the interaction takes place in the contracting branch of the hyperbola. This will lead to a $\chi$ particle with positive momentum $p>0$, as in (b). Therefore, a strong asymmetry in the momentum distribution of the products towards negative momenta can be interpreted as evidence that the detector encounters the expanding branch much more often than the contracting branch, consistent with option (A) in the introduction.\label{three}}
\end{center}
\end{figure}

To investigate (A) and (B), the model detector was chosen to be a particle of a second charged field $\psi$ (see Fig. \ref{three}), interacting with $\phi$ through the vertex  
\begin{equation}
g (\phi\psi^*\chi +{\rm  h.c.}), \label{vertex}
\end{equation}
where $g$ is a coupling constant.
Through this interaction, the $\psi$ particle can anihilate the $\phi$ antiparticle in the pair, producing a neutral particle $\chi$. The kinematics of this process is such that the $\phi$ antiparticle has at most two chances, along its hyperbolic trajectory, of interacting with the detector. The reason is that the center of mass energy of the collision between $\phi^*$ and the detector particle $\psi$ has to be equal to the rest mass $m_\chi$ of the product. This selects the magnitude of the momentum of $\phi$ relative to $\psi$, but we have two options for its sign. 

If the pair nucleates in the rest frame of the detector, as in the left panel of Fig. \ref{three}, then the collision will take place in the expanding branch of the hyperbola, and the momentum of the decay product $\chi$ will be negative $p<0$. On the other hand, if the pair nucleates in a frame which is highly boosted with respect to the detector, then there is a good chance that the interaction may take place in the contracting branch of the hyperbola. This will lead to a $\chi$ particle with positive momentum $p>0$, as in the right panel in Fig. \ref{three}. The  results of Ref.\cite{GKSSV} showed a strong asymmetry in the momentum distribution of the decay products, towards negative momenta. This was interpreted as evidence that the detector encounters the expanding branch much more often than the contracting branch, consistent with option (A). It was therefore concluded that nucleation takes place preferentially in the rest frame of the detector.

The analysis of Ref. \cite{GKSSV} was restricted to a range 
of parameter space where the momentum of the $\phi$ antiparticle at the time of collision is highly relativistic. In this sense, the frame of nucleation was probed rather imprecisely, with a tolerance much larger than the size of the critical bubble. This still left some room for option (B) to be realized.

The purpose of this paper is to sharpen the discussion given in Ref. \cite{GKSSV}, by tightening the precision 
to which the rest frame of bubble nucleation can be determined. 

We start in Section 2 by investigating the timescales which are relevant to bubble nucleation. These include the time $\tau_q$ of quantum fuzziness after nucleation, the size of the critical bubble $r_0$, the timescale $\tau_{nuc}$ that it takes for a small quantum fluctuation in the false vacuum to tunnel into a critical bubble, the time $t_0$ ellapsed since the false vacuum was prepared,
and the average lifetime $\tau_{vac}$ of the false vacuum. The hierarchy between these scales as a function of the instanton action $S_E$ will be clarified.

In Section 3 we briefly review pair production by an electric field in (1+1)-dimensions.
In Subsections 4.1 and 4.2 we consider the response of the model detector (\ref{vertex}), extending the analysis of Ref. \cite{GKSSV} to the range of parameter space where the momentum of the $\phi$ antiparticle at the moment of collision is only mildly relativistic or non-relativistic. {This will allow us to probe the frame of nucleation on scales smaller than $r_0$ and down to the minimum resolution scale $\Delta t\sim \tau_q$. Beyond that, the semiclassical picture breaks down.  In Subsection 4.3 we use a model based on a four-point interaction, where the detector particle is not charged. In this case, we are able to show that even if the interaction takes place within the time interval $\Delta t \sim \tau_q$ in the vicinity of the turning point, there is still a very strong asymmetry in the momentum transferred from the nucleated particles to the detector, in the direction of expansion after the turning point. Our conclusions are summarized in Section 5.}

\section{Timescales}

In the thin wall limit, the action for a vacuum bubble is given by
\begin{equation}
S=-\int M(r) \sqrt{1-\dot r^2}\ dt + \epsilon \int V(r)\ dt. \label{action}
\end{equation}
Here, $M(r)$ is the mass of the domain wall of radius $r$, $\epsilon$ is the difference in energy density between the false and the true vacuum, 
and $V(r)$ is the volume inside the bubble. In $D$ spacetime dimensions, we have
\begin{equation}
M(r) = \sigma\ \Omega\ r^{D-2}, \quad V(r) = \frac{\Omega}{D-1}\ r^{D-1},
\end{equation}
where $\sigma$ is the wall tension, and $\Omega$ is the surface area of the unit $(D-2)$-sphere. A vacuum bubble has zero energy (relative to the false vacuum configuration without the bubble),
and hence it satisfies
\begin{equation}
\sqrt{p_r^2 + M^2} - \epsilon\ V(r) =0. \label{zero}
\end{equation}
Here the radial momentum is given by
\begin{equation}
p_r = \gamma M \dot r, \label{pr}
\end{equation}
with $\gamma=(1-\dot r^2)^{-1/2}$. Up to temporal shifts, the solution of (\ref{zero}) is given by Eq.~(\ref{trajectory}), with
\begin{equation}
r_0 = {(D-1) \sigma\over \epsilon}.  \label{r0}
\end{equation}
Eq.~(\ref{r0}) gives the size of the critical bubble, which is also the acceleration timescale, which is needed for the bubble wall to become relativistic
\begin{equation}
\tau_{acc} \sim r_0.
\end{equation}
To relate $r_0$ to other relevant scales, we introduce the dimensionless combination
\begin{equation} 
\lambda \equiv {\sigma^D\over \epsilon^{D-1}} \sim S_E \gg1.
\end{equation}
Up to a numerical coefficient of order one, this coincides with the instanton action
$S_E$, which can be calculated by substituting the Euclidean  version of (\ref{trajectory}) into the Euclidean version of (\ref{action}). The semiclassical approach to tunneling is only valid when the last strong inequality is satisfied. This will make $\lambda$ a useful expansion parameter.

Eq.~(\ref{zero}) can be thought of as the classical limit of a Schr\"odinger equation for the wave function of the bubble.
This was used in \cite{VKO} to estimate the probability that a bubble of vanishing size may tunnel to the critical size. 
The estimate in \cite{VKO} is in qualitative agreement with the decay rate per unit volume $\Gamma \sim e^{-S_E}$ which is obtained by the instanton methods \cite{Coleman}. The lifetime $\tau_{vac}$ for the false vacuum is therefore exponential in $\lambda$,
\begin{equation}
\ln \tau_{vac} \propto \lambda.
\end{equation}

As mentioned in the introduction, the semiclassical description is not adequate at the classical turning point $r=r_0$, where the radial momentum $p_r$ vanishes. However, it does become very accurate for $t\gg \tau_q$, where $\tau_q$ can be determined from the condition 
$[r(\tau_q)-r_0]\ p_r(\tau_q) \sim 1$. This leads to 
\begin{equation}
\tau_q \sim \lambda^{-1/3} r_0 \ll r_0.
\end{equation}
Hence, the intrinsic uncertainty in the time of nucleation is much smaller than the bubble size
(as illustrated in Figs. \ref{bubble} and \ref{boost}).

The uncertainty $\tau_q$ should not be confused with the timescale $\tau_{nuc}$ that it takes for the critical bubble to 
form out of a tiny false vacuum fluctuation. In the semiclassical picture, it is somewhat unclear how to characterize 
$\tau_{nuc}$. Causality suggests that this time should at least be as large as the size of the critical bubble,
\begin{equation}
\tau_{nuc} \gtrsim r_0.
\end{equation}
The scale $\tau_{nuc}$ will be characterized more precisely in the next Section, where we consider pair production as a model for vacuum decay in (1+1) dimensions. 

In summary, we are led to the following hierarchy of scales,
\begin{equation}
\tau_q \ll \ r_0 \lesssim \tau_{nuc}\lesssim t_0 \ll \tau_{vac}.   \label{scales}
\end{equation}
where the strong inequalities are parametrically enhanced the larger is $\lambda \sim S_E$.
In (\ref{scales}), $t_0$ is the time ellapsed from the hypersurface of initial conditions to the moment when the 
bubble nucleates. This should be at least marginally larger than $\tau_{nuc}$. We impose the last strong inequality on $t_0$ because our primary interest is in bubbles which nucleate in isolation, without interference from collisions with other bubbles. Bubbles that form at $t_0\ll \tau_{vac}$ are very likely to remain isolated for a period of time which is much larger than all other scales involved in the problem. Finally, it should be noted that $\tau_{vac}$ grows exponentially with $\lambda$, whereas for all other scales the dependence is 
power law. In this sense, we can practically think of $\tau_{vac}$ as infinite. 

In the (1+1)-dimensional example which we consider in the following Section, the electric field producing the pairs is external and the pairs do not interact with each other. In this idealized situation, the scale $\tau_{vac}$  does not play any role, and the relevant parameters in the above discussion are given by
\begin{equation}
\sigma = m,\quad\quad \epsilon= eE, \quad \quad \lambda= {m^2 \over eE},
\end{equation}
where $m$ is the mass of the charged particles that nucleate, $e$ is their charge and $E$ is the electric field.

\section{Pair production}
\label{spp}

Here we briefly review Schwinger pair production, which serves as a model for bubble nucleation in (1+1) dimensions. The main advantage is that the nucleated pairs are treated fully quantum mechanically. In the following Sections, we shall consider the detection of the nucleated pairs. Our conventions will follow those of Ref. \cite{GKSSV}.

Consider a charged scalar field $\phi$, coupled to an external electric field with gauge potential 
\begin{equation}
A_\mu = \delta^x_\mu A(t). \label{gauge} 
\end{equation}
By spatial homogeneity, we can separate $\phi(x)$ into Fourier modes, which satisfy the equation. 
\begin{equation}
\ddot \phi_k + w_k^2 \phi_k =0, \label{modeq}
\end{equation}
where 
\begin{equation}
w_k^2 = m^2 + (k - e A)^2.\label{waka}
\end{equation}
Here, $m$ is the mass of the field $\phi$ and $e$ is its charge. 

For $A(t) = - E t$, we have a constant electric field $E$, and (\ref{modeq}) becomes 
\begin{equation}
\phi_k'' + (\lambda + z^2) \phi_k = 0, \label{parcyl}
\end{equation}
where
\begin{equation}
z\equiv \sqrt{eE}(t+k/eE), \quad\quad \lambda\equiv {m^2\over eE},\label{inra}
\end{equation}
and primes indicate derivatives with respect to $z$.
Eq.~(\ref{parcyl}) has normalized positive frequency solutions
\begin{equation}
\phi_k = \frac{1}{(2eE)^{1/4}}\, e^{i\frac{\pi}{4}\nu^*}D_{\nu^*} [-\sqrt{2}\ e^{-i\pi/4}  z],
\label{inr}
\end{equation} 
where 
\begin{equation}
\nu=-{1+i \lambda \over 2}.
\end{equation}
From the asymptotic expansion of the parabolic cylinder function at large negative $z$, we have
\begin{equation}
\phi_k \approx \frac{1}{(2eE)^{1/4}} \left(-\sqrt{2}\,z\right)^{\nu^* }e^{\frac{i}{2}z^2}\,, 
\qquad{\rm for}\quad
-z \gg  |\nu| \,. \label{asym}
\end{equation} 
Therefore, leaving aside an irrelevant phase, we have
\begin{equation}
\phi_k \approx (2w_k)^{-1/2} e^{-i\int^t w_k dt'} \quad (t \to -\infty). \label{pve}
\end{equation}
The ``in" positive frequency mode $\phi_k$ can be 
expressed as 
\begin{eqnarray}
\phi_k = \alpha_k\,\phi_k^{\rm out} + \beta_k\,\phi_k^{{\rm out}*}\,,
\label{Bogoliubov}
\end{eqnarray}
where  $ \phi_k^{\rm out}(z)
= (2eE)^{-1/4}\, e^{-i\frac{\pi}{4}\nu}\,D_{\nu} [(1+i)  z]$
is the positive frequency ``out" mode, which behaves as 
the right hand side of (\ref{pve}) at late times, $t\to +\infty$.

\begin{figure}
\begin{center}
\vspace{-2cm}
\includegraphics[width=14cm]{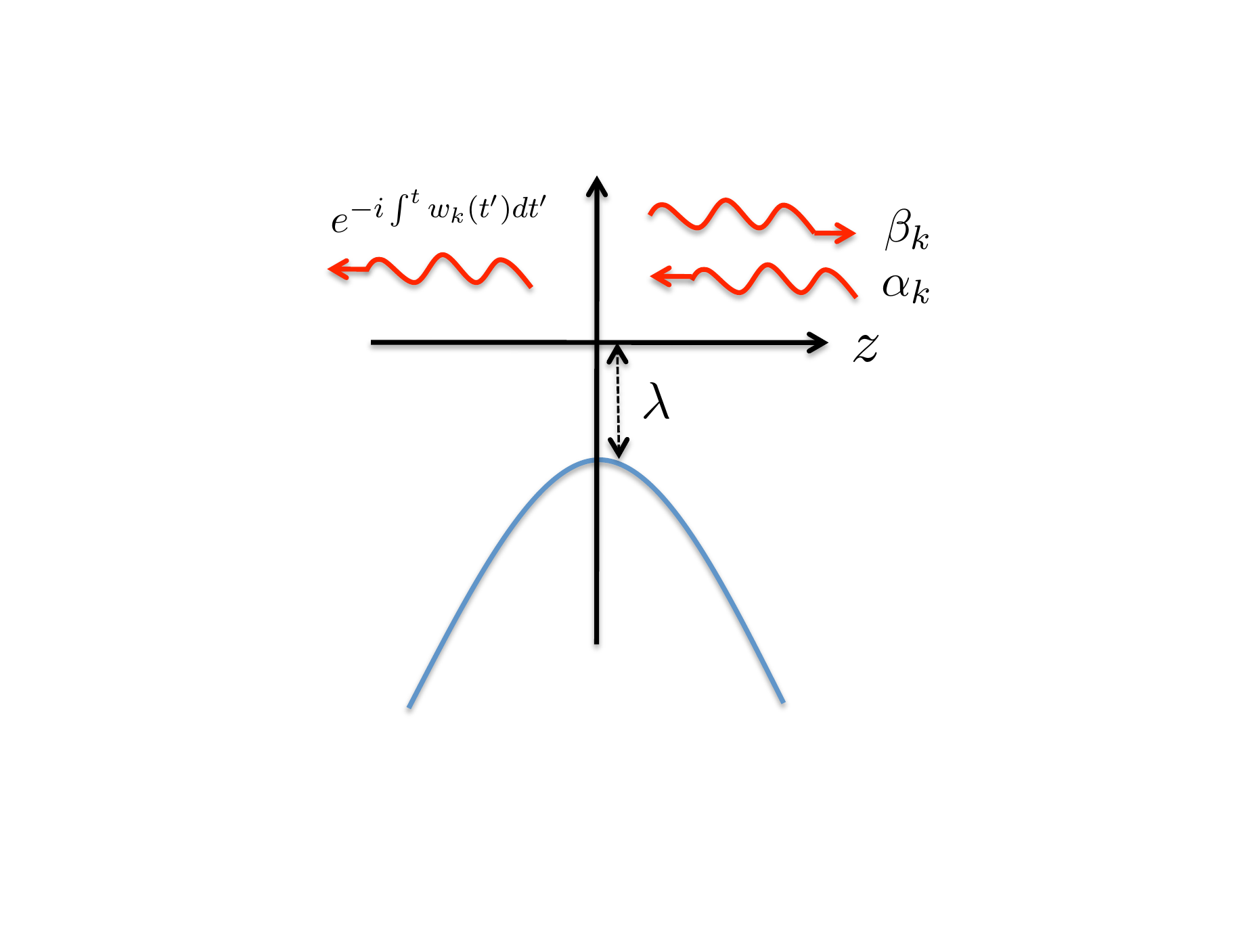}
\vspace{-1.5cm}
\caption{The effective potential in the Schr\"odinger Eq.~(\ref{parcyl}). The scattering which determines the
Bogolubov coefficients is overbarrier, so the reflection coefficient $\beta_k$ is exponentially suppressed.
 \label{harmonic}}
\end{center}
\end{figure}

The number density of ``out'' particles in the ``in'' vacuum per unit momentum space interval 
is given by 
\begin{equation}
{dn\over dk} = {1\over 2\pi} |\beta_k|^2.  \label{dndk}
\end{equation}
Eq.~(\ref{parcyl}) can be thought of as a Schr\"odinger
equation with potential $V_{eff}=-(\lambda +z^2)$, and the determination of the Bogolubov
coefficients $\alpha_k$ and $\beta_k$ amounts to solving a scattering problem in this potential 
(see Fig. \ref{harmonic}).  Note that $k$ enters Eq.~(\ref{parcyl}) only through the combination $k +eEt$.
This means that the scattering near $z=0$ occurs approximately at the time 
$t \approx -k/(eE)$, but it is otherwise independent of $k$.
Eq.~(\ref{dndk}) can then be rewritten as
\begin{equation}
\Gamma= dn /dt = {eE |\beta_k|^2/(2\pi)}, \label{ratebeta}
\end{equation}
where the coefficient $\beta_k=e^{i\pi\nu^*} $ can be easily calculated from a standard linear relation between the parabolic cylinder functions appearing in $\phi_k$ and $\phi^{\rm out}_k$ (see e.g. \cite{GKSSV} and references therein). This 
leads to Schwinger rate
\begin{equation}
\Gamma = {eE\over 2\pi} e^{-\pi \lambda}, \label{schw}
\end{equation}
The exponent in (\ref{schw}) coincides with the Euclidean action $S_E=\pi \lambda$, in agreement with the instanton approach, which is valid for large $S_E$.

The asymptotic expansion (\ref{asym}) for  $\phi_k$, and the analogous one for 
$\phi_k^{out}$, are valid for $|z|\gg |\nu|$. Hence the timescale of scattering off the potential in Fig. \ref{harmonic}, where the  mixing of positive and negative frequency modes occurs, is at most of order 
\begin{equation}
\tau_{mixing} \lesssim \nu (eE)^{-1/2} \sim  \lambda^{1/2} r_0. 
\end{equation}
On the other hand, this seems to be a rather crude upper bound. Mode mixing will be at its peak when the non-adiabaticity parameter 
\begin{equation}
f_k(t)=\left|{\dot w_k \over w_k^2}\right|
\end{equation}
is at its maximum. The function $f_k$ is symmetric around the time $t=t_k \equiv k/eE$ (where it vanishes) and has two peaks which are at a distance $\Delta t \sim r_0$ away from $t_k$. Hence, we can estimate 
\begin{equation}
\tau_{mixing} \sim r_0. 
\end{equation}
Identifying this with the semiclassical 
timescale $\tau_{nuc}$ for nucleation of a pair out of a vacuum fluctuation suggests that
\begin{equation}
\tau_{nuc} \sim r_0.  \label{taunuc}
\end{equation}
This is consistent with the estimate $\tau_{nuc} \gtrsim r_0$ which we mentioned in the previous subsection.

Stronger evidence for (\ref{taunuc}) can be found by considering the case of an electric field $\hat E(t)$ which is turned on and off on a timescale $t_0$:
\begin{equation}
\hat E(t) = {E \over \cosh^2(t/t_0)}. 
\end{equation}
The corresponding gauge potential is given by $A(t)= -E t_0 \tanh(t/t_0)$, and the mode equation (\ref{modeq}-\ref{waka}), can in this case be solved 
in terms of hypergeometric functions . The Bogoliubov coefficient is given by (see e.g. \cite{Gavrilov:1996pz} 
and references therein)
\begin{equation} 
|\beta_k|^2 = {\cosh^2\left[\pi \sqrt{(eEt_0^2)^2-{1\over 4}}\right] +\sinh^2\left [{\pi t_0\over 2} (w_+-w_-)\right]
\over \sinh(\pi t_0 w_+) \sinh(\pi t_0 w_-)}, \label{beta}
\end{equation}
where $w_\pm = \sqrt{m^2+(k\mp eE t_0)^2}$. For the mode with conserved momentum $k=0$, the physical momentum $k_{phys} = k - e A(t)$ vanishes precisely at the time $t=0$, which therefore corresponds to the 
turning point. For this mode, $w_+=w_-$ and (\ref{beta}) simplifies to 
\begin{equation} 
|\beta_{k=0}|^2 = {\cosh^2\left[\pi \sqrt{(eEt_0^2)^2-{1\over 4}}\right]
\over \sinh^2 \left[\pi t_0 \sqrt{m^2 +(eEt_0)^2}\right]}. \label{beta0}
\end{equation}
This expression determines, through Eq.~(\ref{ratebeta}), the rate of pair production
at the time when the electric field is at its maximum $\hat E(t=0) =E$. Further simplification
occurs when the timescale $t_0$ during which the electric field is switched on is bigger than 
the scale of quantum fuzziness $t_0\gg \tau_q \sim \lambda^{-1/3} r_0$. In this case, we have 
\begin{equation}
eE t_0^2  > \lambda^{1/3} \gg 1. \quad\quad    (\tau_q < t_0).
\end{equation}
Now, we can distinguish two cases. For $t_0\ll r_0$, we have $eEt_0 \ll m$, which leads to 
\begin{equation}
|\beta_{k=0}|^2 \approx e^{-2\pi m t_0}, \quad\quad (\tau_q<t_0 \ll r_0), \label{switch}
\end{equation}
while for $t_0\gg r_0$, we have $eEt_0 \gg m$, and
\begin{equation}
|\beta_{k=0}|^2 \approx e^{-\pi\lambda}. \quad\quad (t_0 \gg r_0). \label{adiabatic}
\end{equation}
In the first case, Eq.~(\ref{switch}), the pair production rate near $t=0$ is independent of the electric field.
Clearly, this is unrelated to the Schwinger process, and corresponds to particle creation by the non-adiabaticity of the switching process. On the other hand, we see from Eq.~(\ref{adiabatic}) that if the electric field changes
on timescales which are larger than the radius of the critical bubble, then we obtain the adiabatic Schwinger 
pair production rate. This strongly suggests that the timescale of nucleation of a pair from a small vacuum 
fluctuation satisfies $\tau_{nuc} \lesssim r_0$. In combination with the bound $\tau_{nuc} \gtrsim r_0$ which is suggested by causality, we are again lead to the estimate (\ref{taunuc}).  In conclusion, out of the scales in (\ref{scales}), two of them are of comparable size $r_0 \sim \tau_{nuc}$.

\section{Charged detector (3-point interaction)}

Let us start by reviewing the model detector which was introduced in Ref. \cite{GKSSV}. This involves a charged $\psi$ particle interacting with the antiparticle of a nucleated pair of the charged field $\phi$, through the 
vertex
\begin{equation}
g (\phi\psi^*\chi +{\rm  h.c.}), \label{vertex2}
\end{equation}
In the gauge (\ref{gauge}), the canonical momentum of charged particles is conserved. Denoting by $-k$ the momentum of the $\phi$ antiparticle, by $q$ the momentum of the $\psi$ particle, and by $p$ the momentum of the $\chi$ particle, we have
\begin{equation}
p=q-k. \label{pqk}
\end{equation}
The physical momentum of charged particles changes with time, due to the acceleration caused by the electric field,
\begin{equation}
q_{phys} = q+eEt, \quad\quad k_{phys} = k+eEt. \label{qphys}
\end{equation}
From (\ref{pqk}) and (\ref{qphys}), we also have $p=q_{phys}-k_{phys}$. 

Note that the $\psi$ particle detector is accelerated by the electric field. At early times it is highly relativistic with negative 
velocity, and at late times it is highly relativistic with positive velocity, following a hyperbolic trajectory. As a consequence, the momentum distribution 
of the resulting $\chi$ particles is actually Lorentz invariant \cite{GKSSV}:
\begin{equation}
dN_\chi = C {dp \over w_\chi}, \label{li}
\end{equation}
where  $N_\chi$ is the number of $\chi$ particles, $p$ is their spatial momentum, 
\begin{equation}
w_\chi=\sqrt{p^2 + m_\chi^2}, \label{wp}
\end{equation}
and $C$ is a constant.
The form of (\ref{li}) can be understood as follows. The interaction of $\psi$ with nucleated pairs has a constant probability of occurring per unit proper time interval $d\tau$ along the trajectory of the $\psi$ particle. This trajectory can also be thought of as an orbit of the Lorentz boosts, with parameter 
$\phi_v=a_\psi \tau$, where $a_\psi = eE/m_\psi$ is the proper acceleration of the detector particle. If we consider two collisions which look identical in the detector's rest frame, separated by a time delay $d\tau$, the corresponding 
momenta $p$ will be related by a boost with parameter $d\phi_v = (dp/w_\chi) =  a_\psi d\tau$.  Hence, Eq.~(\ref{li}) simply expresses the 
fact that the probability of producing a $\chi$ particle per unit proper time is constant. An immediate consequence is that the distribution (\ref{li}) cannot inform us about the frame of nucleation.

This difficulty is avoided by switching on the interaction (\ref{vertex}) for a short period of time $T$. This can be implemented by substituting the coupling $g$ in (\ref{vertex2}) by a time dependent coupling with a Gaussian profile  \cite{GKSSV}:
\begin{equation}
g\to g(t) = g\ e^{-t^2/T^2}.
\end{equation}
If we choose
\begin{equation}
q=0, 
\end{equation}
then, during the time interval $|t| \lesssim T$, we have 
\begin{equation}
|q_{phys}| \lesssim eET \ll m_\psi<m_\chi-m_\phi. \label{nonrel}
\end{equation}
For any time interval $T$, we may choose the mass $m_\psi$ of the detector to be heavy enough, so that the above strong inequality holds, and in this case, the detector is essentially
at rest within the time interval $T$ before the interaction. The last inequality in  (\ref{nonrel}) indicates that we are choosing the mass $m_\chi$ of the decay product to be larger than $m_\psi+m_\phi$. Otherwise the detection of $\phi$ antiparticles would be forbidden by energy conservation (at least in the absence of an electric field and of any time dependence in the coupling $g$).
Within this setup, a significant asymmetry in the momentum distribution for the decay products $\chi$ towards negative values of $p$ will indicate that the detector is more likely to be hit from the right, indicating that the interaction is more likely to happen in the ``expanding" branch of the hyperbola, when the $\phi$ antiparticle is moving to the left. On the other hand, a more symmetric distribution would indicate that there is a significant chance of detecting the anti-particle also in the ``contracting" branch (see Fig. \ref{three}) .  

\begin{figure}
\begin{center}
\vspace{-2cm}
\includegraphics[width=14cm]{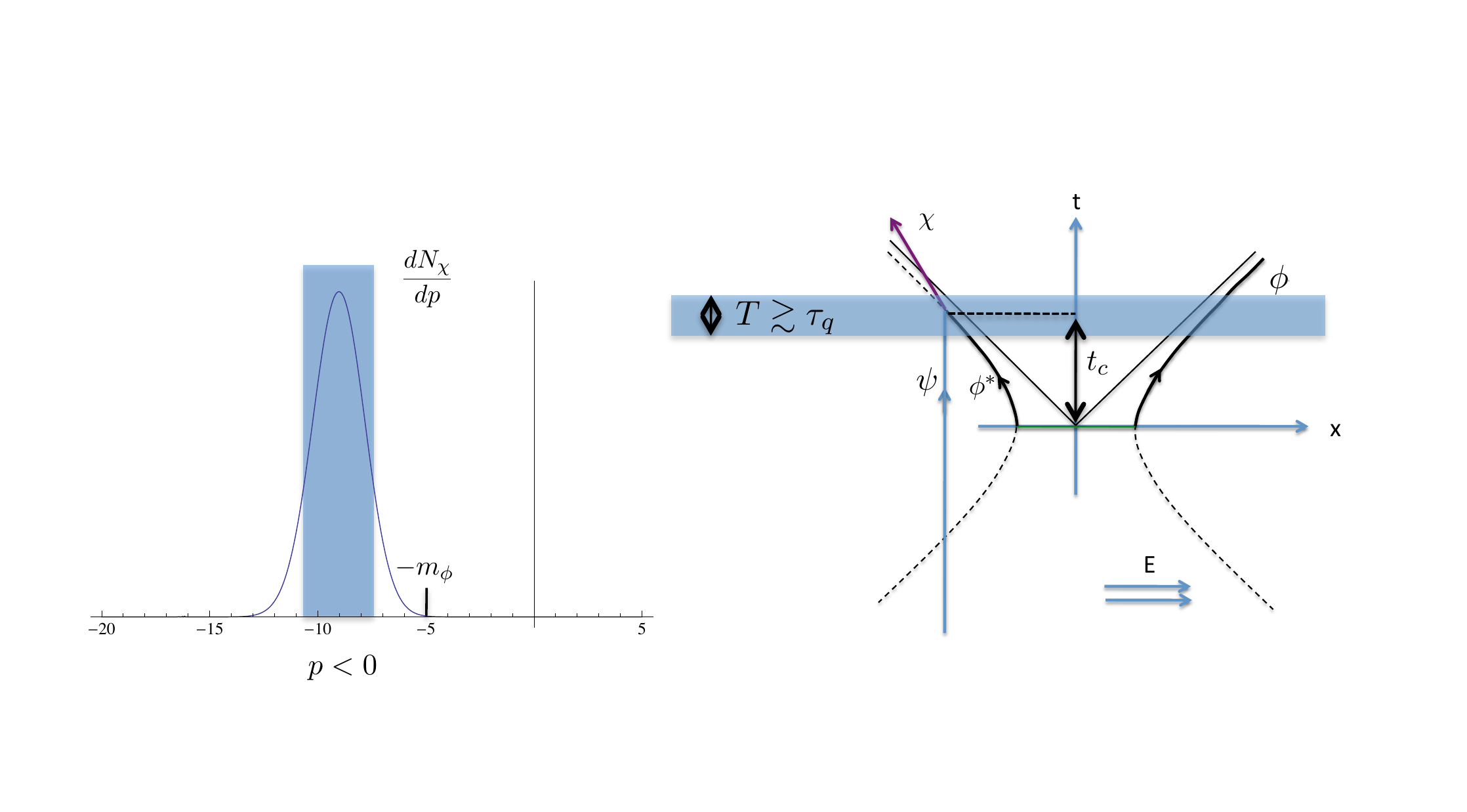}
\vspace{-1.5cm}
\caption{The left panel shows the momentum distribution of $\chi$ particles, Eq.~(\ref{big}), for $m_\phi = 5\ (eE)^{1/2}$ (corresponding to 
$\lambda=25$), with $m_\chi=110\ (eE)^{1/2}$, $m_\psi = 100\ (eE)^{1/2}$ and $\alpha=eET^2 =\lambda^{1/3}$. 
The width $\Delta p$ of the distribution can be estimated from (\ref{nonrel}). During the time interval of duration $T$ 
when the interaction is switched on, indicated by shading in the right panel, the detector particle is accelerating
and its momentum changes by the amount$\Delta q_{phys} \sim eET$. This leads to a similar spread 
$\Delta p \sim eET$ in the momentum distribution of the products. The distribution is clearly biased towards negative 
momenta $p<0$. This means that the detector $\psi$ never encounters the $\phi$ antiparticle in the contractig branch
of the hyperbola (represented by the dotted line). \label{dist}}
\end{center}
\end{figure}

The momentum distribution of $\chi$ particles for an interaction of finite duration $T$ was calculated in \cite{GKSSV}, and it is given by
\begin{equation}
{dN_\chi \over dp} = {1\over 2\pi} \left|{\cal A}(p,q=0,T)\right|^2, \label{distribution}
\end{equation}
where 
\begin{equation}
{\cal A}(p,q=0,T) \equiv \int_{-\infty}^{\infty} dt\ g(t)\ \psi_{q=0}(t)\ \phi^*_{-p}(t)\  \chi^*_p(t). \label{amplitude}
\end{equation}
Here, the mode function $\phi_{k}$ is given by (\ref{inr}), and similarly for $\psi_q$ (with the mass $m_\phi$ replaced with $m_\psi$). The mode function for 
the neutral particle is simply given by $\chi_p = (2w_\chi)^{-1/2} e^{-i w_\chi t}$. It follows from Eq.~(\ref{nonrel}) that the $\psi$ particle stays non-relativistic throughout the interaction, so we can approximate $\psi_0 \approx (2 m_\psi)^{-1/2} e^{-1/2} e^{-im_\psi t}$. Here, particle production of the $\psi$ field is also neglected. Within this approximation, the amplitude (\ref{amplitude}) can be calculated 
explicitly \cite{GKSSV}, and we have
\begin{equation} 
{dN_\chi \over dp}  \approx  {C_T \over w_\chi} \ \exp\left[- 2\frac{\alpha(p^2+W^2)}{(4+\alpha^2)eE}\right]  \left|D_{\nu} \left( Z\right)\right|^2 ,
\label{big}
\end{equation}
where 
\begin{equation}
\nu=-(1+i\lambda)/2 \label{nuind}
\end{equation}
and
\begin{equation}
W \equiv w_\chi -m_\psi. \label{deltaw}
\end{equation}
Note that if energy conservation were exact, then $W$ would be the energy of the charged antiparticle $\phi$.
In (\ref{big}) we have also introduced the parameter
\begin{equation}
\alpha \equiv eET^2,
\end{equation}
and the variable
\begin{equation}
Z \equiv \sqrt{2\over ieE}\  {\alpha W + 2ip \over \sqrt{4+\alpha^2}}. \label{zdef}
\end{equation}
Finally, the momentum independent prefactor in (\ref{big}) is given by
\begin{equation} 
C_T=\frac{g^2\  e^{-3\pi \lambda/4}}{2 m_\psi (2eE)^{3/2}}  {\alpha \over \sqrt{4+\alpha^2}}\ e^{\lambda \tan^{-1}(2/\alpha)}.
\end{equation}
A plot of the distribution (\ref{distribution}) is given in the left panel of Fig. \ref{dist}, for $m_\phi = 5\ (eE)^{1/2}$ (corresponding to  $\lambda=25$), with $m_\chi=110\ (eE)^{1/2}$, $m_\psi = 100\ (eE)^{1/2}$ and $\alpha=\lambda^{1/3}$. It is clear that the momentum distribution is highly biased towards negative momenta, $p<0$.

Denoting by $t_c$ the time of collision minus the time at which nucleation occurs,
the physical momentum of the $\phi$ antiparticle at the time of collision is given by $p= - eE t_c$, and we have
\begin{equation}
{p\over m_\phi}= - {t_c \over r_0}.\label{ratios}
\end{equation}
Consider, for illustration, the distribution in Fig \ref{dist}. This is peaked at $p^2 \sim m^2_\phi$, which corresponds to a time of collision  $|t_c| \sim r_0$ (see the right panel of Fig. \ref{dist}). For the chosen values of the masses $m_\phi, m_\psi$ and $m_\chi$,  both signs of $p$ and $t_c$ would be kinematically allowed, but we find that detection only occurs in the expanding branch of the hyperbola. From this we can conclude that the boost parameter $\phi_v$ which relates the frame of nucleation and the frame of the detector (see Fig. \ref{boost}) is bounded by 
\begin{equation}
\sinh |\phi_v| \lesssim {t_c \over r_0},\label{preci}
\end{equation}
This sets the accuracy to which the frame of nucleation is determined by this particular interaction. 
In the example represented in Fig. \ref{dist},  we have $t_c \sim r_0$, which leads to $|\phi_v| \lesssim 1$. This is not too accurate, but we shall see that the precision can be increased parametrically 
in $\lambda$, by choosing $m_\chi-m_\psi$ to be sufficiently close to the rest mass of the antiparticle $\phi$.

Our goal here is to examine the regime where both the time of collision $t_c$ and the time interval 
$T$ during which the interaction is 
switched on are marginally bigger than, but very close to the scale $\tau_q$, 
\begin{equation}
t_c \sim T \sim {\rm (few\ times)}\  \tau_q \ll r_0.
\end{equation}
In this way, we can probe the time of nucleation to the finest possible resolution consistent with the uncertainty principle. 
To realize this situation, we need to explore the regime of very large $\lambda$. The reason is that the hierarchy between the quantum scale and the size of the critical bubble, $\tau_q \sim \lambda^{-1/3} r_0$, is only appreciable for, say, $\lambda \gtrsim 10^3$. In this parametric range, the order of 
the parabolic cylinder function in (\ref{big}) is very large $|\nu|\gtrsim 10^3$. As a result, a numerical study becomes quite cumbersome, and an analytic approach will prove more illuminating.

{First, we note that from (\ref{nonrel}), and (\ref{deltaw}) we have
\begin{equation}
W>m_\phi,\label{deltawbig}
\end{equation}
and using (\ref{zdef}), we have}
\begin{equation}
|Z| \gtrsim {W\over \sqrt{eE}}\gtrsim \lambda^{1/2} \gg 1.
\end{equation}
The parabolic cylinder functions  of large order $|\nu|\sim \lambda \gg1$ and argument $Z\gtrsim \lambda^{1/2}$ can
be estimated by using an integral representation of $D_\nu(Z)$, and then evaluating this integral by using 
the method of steepest descents. This is done in Appendix \ref{appA}.  {In this approximation, the distribution (\ref{big}) is given by}
\begin{equation}
{dN_\chi }  \approx  {g^2\over 8 eE\ m_\psi}\ {e^{-\pi\lambda}\over   P}\ \exp\left[{- 2\frac{(p+P)^2}{(eET)^2}}\right]{dp\over w_\chi}.
\label{small}
\end{equation}
Here, we have introduced
\begin{equation}
P\equiv (W^2-m_\phi^2)^{1/2}.
\end{equation}
Note from Eq.~(\ref{deltaw}) that $W=w_\chi-m_\psi$. Since, from (\ref{nonrel}) $\psi$ is assumed to be non-relativistic, 
$W$ is the difference in energy between the $\chi$ and the $\psi$ particle. Hence, if energy is approximately conserved,
we have $W \approx w_\phi \equiv (p^2+m_\phi^2)^{1/2}$, and so we expect $|p| \approx P$, in agreement with (\ref{small}). 
{Even though $P$ is strictly speaking $p$ dependent, it will be approximately constant for $p\ll m_\chi$. The mass of the $\chi$ particle can be taken as large as desired, so we can always think of 
\begin{equation}
P \approx m_\phi \sqrt {\left({m_\chi-m_\psi \over m_\phi}\right)^2 -1}, \label{massP}
\end{equation}
as a constant parameter in the momentum range of interest.
In deriving (\ref{small}) we have also assumed, according to Eqs. (\ref{minP}) and (\ref{minP2}), that 
\begin{equation}
P \gtrsim \lambda^{-1/3} m_\phi,
\label{plarge}
\end{equation}
and
\begin{equation}
P \gg \alpha^{-1} m_\phi, \label{bal}
\end{equation}
with
\begin{equation}
\alpha = eE T^2 \gg 1.
\end{equation}
Eq.~(\ref{small}) is the main result of this Section.} 

The momentum distribution (\ref{small}) is peaked at negative values
\begin{equation}
p\approx -P <0, 
\end{equation}
with a width 
\begin{equation}
\Delta p \sim eET. \label{width}
\end{equation}
The origin of the spread in momentum in Eq.~(\ref{width}) is simple to understand: 
it corresponds to the variation in the momentum of 
a charged particle (the detector) due to the force exerted by the electric field $E$ during the time interval $T$.

{We are particularly interested in small values of the momentum, for which (\ref{plarge}) is marginally satisfied,
\begin{equation}
|P| \sim \lambda^{-1/3} m_\phi.  \label{marginal}
\end{equation}
In this case, the condition (\ref{bal}) requires $\alpha \gg \lambda^{1/3}$, or
\begin{equation}
T\gtrsim \lambda^{-1/3} r_0 \sim \tau_q.
\end{equation}
On the other hand, the asymmetry in the momentum distribution will only be appreciable provided that 
$|P| \sim \lambda^{-1/3} m_\phi \gtrsim \Delta p \sim eET$. This requires
\begin{equation}
T\lesssim \lambda^{-1/3} r_0 \sim \tau_q.
\end{equation}
Hence, we can have a well defined peak at $p \sim - \lambda^{-1/3} m_\phi$ by choosing the mass difference $m_\chi-m_\psi$ in (\ref{massP}) so that $P$ satisfies (\ref{marginal}), and by using $T \sim \tau_q$. 
Combining with  
(\ref{ratios}) and (\ref{preci}), we find that the boost parameter $\phi_v$ which relates the rest frame of the detector with the rest frame of bubble nucleation can be parametrically bounded with precision
\begin{equation}
|\phi_v| \sim \lambda^{-1/3} \sim S_E^{-1/3}, \label{rfp}
\end{equation}
corresponding to a time of collision $t_c \sim \lambda^{-1/3} r_0 \sim \tau_q$. Hence, the larger the Euclidean action $S_E$, the sharper is the precision to which both rest frames coincide.}

To understand the origin of (\ref{rfp}), we note that energy conservation at the time of collision becomes imprecise in inverse proportion to the time of interaction $\Delta t=T$. 
For non-relativistic particles, we have
\begin{equation}
{(\Delta p)^2 \over 2 m_\phi} \gtrsim (\Delta t)^{-1}. \label{kedt}
\end{equation}
Combining (\ref{width}) and (\ref{kedt}), the minimum uncertainty in the velocity of the $\phi$ particle is given by
\begin{equation}
\Delta v \gtrsim \lambda^{-1/3}, \label{limitation}
\end{equation}
which is saturated by $\Delta t \sim \tau_q$.
{This uncertainty in the velocity of a charged particle embedded in an electric field cannot be reduced.}

Note also that if we use a 3-point interaction such as (\ref{vertex2}), and we take the neutral field $\chi$ as the detector, then the only product of the interaction is a charged particle $\psi$ which accelerates uniformly. Because of that, if the interaction is not switched on and off, the resulting distribution of physical momenta for $\psi$ is Lorentz invariant \cite{GKSSV}, and shows no asymmetry in momentum. Hence, we cannot tell whether the detector has been hit from the right or from the left. The only way to check for this asymmetry would be by considering an interaction of finite duration $T$. But in that case we encounter the same situation which we discussed above: the momentum at the time of collision is uncertain by the amount given in Eq.~(\ref{width}), and the precision in the determination of the frame of nucleation is limited by (\ref{limitation}).

\section{Neutral detector (4-point interaction)}

To avoid {some of the difficulties} associated with the time variation in the momentum of charged particles, in this Section we consider a 4-point interaction
\begin{equation}
\hat g (\chi_1 \chi_2^* \phi \psi^* + \chi_1^*\chi_2 \phi^*\psi), \label{vertex4}
\end{equation} 
where both the detector particle $\chi_1$ and one of the 
{products of interaction} $\chi_2$ are neutral with respect to the electric field. {The main advantage of this setup is that we will not need 
to switch on and off  the interaction in order to find an asymmetry in the momentum distribution of the products.}

\begin{figure}
\begin{center}
\vspace{-2cm}
\includegraphics[width=14cm]{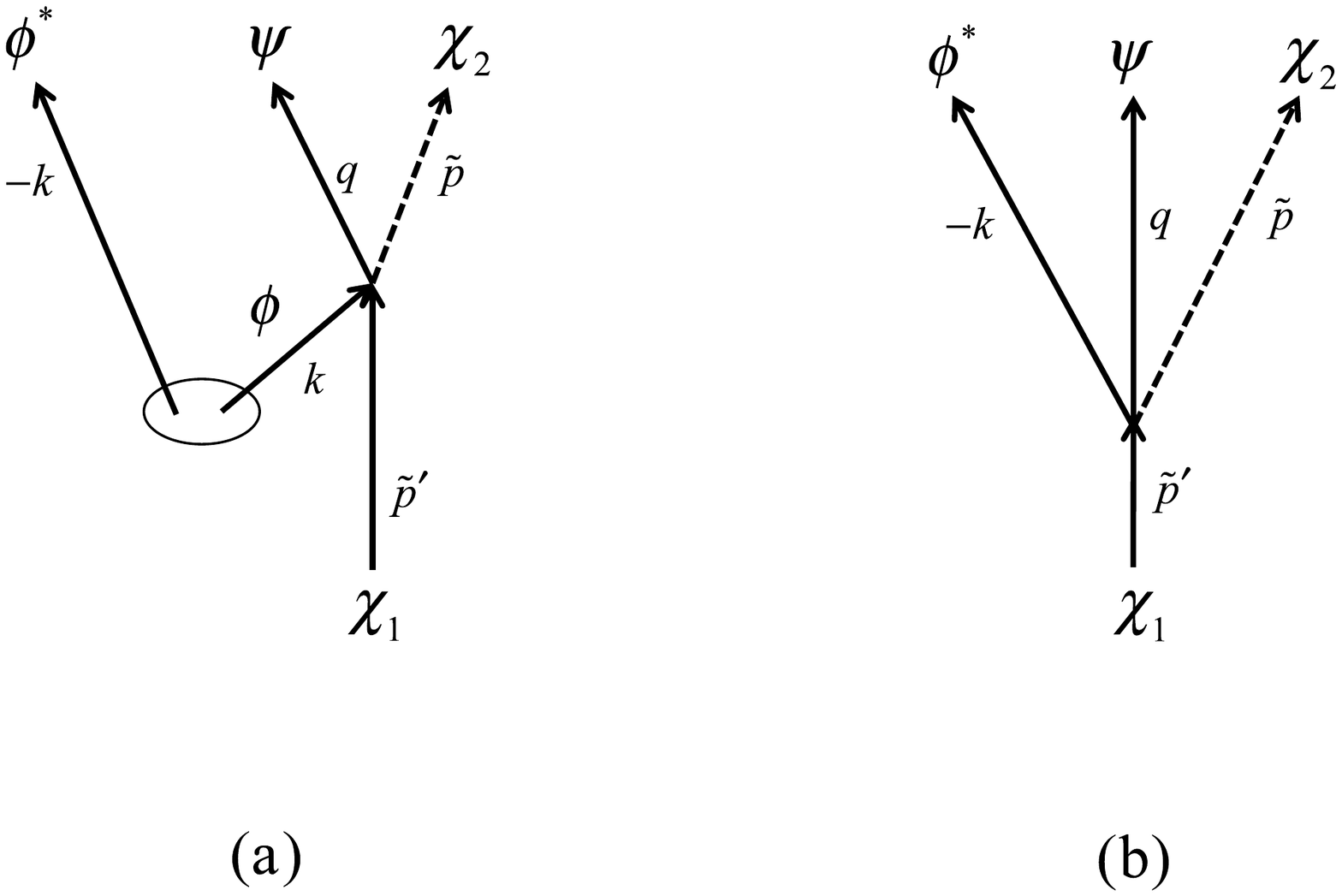}
\vspace{-1.5cm}
\caption{The four-point interaction. Panel $(a)$ represents detection of a $\phi$ particle by interaction with $\chi_1$, 
while panel $(b)$ represents direct decay of $\chi_1$. The latter process is suppressed for $m_1 < m_2 +m_\phi+m_\psi$, which is guaranteed for $m_1=m_2\equiv m_\chi$.
 \label{4pointfig}}
\end{center}
\end{figure}

Even though $\chi_1$ and $\chi_2$ are neutral, we take them to be complex, so that they carry a conserved global charge. The reason is that we want to detect $\phi$ particles, but not $\phi$ anti-particles. With the interaction (\ref{vertex4}) this is achieved by using an incoming $\chi_1$ particle as our detector (as opposed to a $\chi_1$ antiparticle). {For simplicity, we choose 
\begin{equation}
m_1=m_2\equiv m_\chi.
\end{equation}
This automatically suppresses the direct decay of the $\chi_1$ particle into the other three (see Fig. \ref{4pointfig}). }

The expectation value in the number of $\chi_2$ particles can be calculated by using the same techniques which were used in Ref. \cite{GKSSV} 
for the case of the 3-point interaction. This leads to the expression 
\begin{equation}
{dN_2 \over dq dp} = {1\over (2\pi)^2} |{\cal A}_2(\tilde p'=0,\tilde p=p,q)|^2,\label{dist2}
\end{equation}
where the amplitude is given by
\begin{equation}
{\cal A}_2(p,q) = \hat g \int_{-\infty}^{\infty}  dt'\ \phi^*_{k}(t')\  \psi^*_q(t')\ \chi^{}_{1,0}(t')\ \chi^{*}_{2,p}(t'),\label{amplitu2}
\end{equation}
with 
\begin{equation}
k=q+p.
\end{equation}
{Without loss of generality, we have taken the initial $\chi_1$ particle to be at rest, $\tilde p'=0$,
while $\chi_2$ has momentum $\tilde p =p$.}
For the neutral particles $\chi_i$, the mode functions in the amplitude (\ref{amplitu2}) are given by 
\begin{equation}
\chi_{1,0} =(2 m_\chi)^{-1/2} e^{-im_\chi t},\quad \chi_{2,p} =(2 w_{2})^{-1/2} e^{-i w_{2}t}. 
\end{equation}
where
\begin{equation}
w_2=(p^2+m_\chi^2)^{1/2}.
\end{equation}
For the charged particles, the mode functions $\psi_q$ and $\phi_k$ are given by Eq.~(\ref{inr}), 
where the corresponding values of the masses and momenta should be used in the definitions of $z$ and $\lambda$  in
Eq.~(\ref{inra}).

\subsection{Kinematics}

{Since we have chosen $m_1=m_2$, the neutral sector can be thought of as a detector of mass $m_\chi$ which is being hit by a charged particle $\phi$. During the interaction, part of the momentum of this charged particle is transferred to the detector, and part of it to a charged particle $\psi$, whose mass is $m_\psi>m_\phi$.}
{It is convenient to introduce the invariant momentum transfer (which is space-like),
\begin{equation}
U(p^2) \equiv (\tilde p^\mu-\tilde p'^\mu )(\tilde p_{\mu}-\tilde p'_{\mu}) = 2 m_\chi\sqrt{m_\chi^2 + p^2} -2 m_\chi^2>0.\label{L}
\end{equation}
The physical momenta of the charged $\psi$ and $\phi$ particles are given, respectively, by
\begin{equation}
q_{phys} = q+eEt,\quad\quad k_{phys} = k+eE t. \label{phmo}
\end{equation} 
Because of the time dependence in (\ref{phmo}) we cannot properly speak of energy conservation at the collision. Rather, as mentioned around Eq.~(\ref{limitation}), there will be some uncertainty in the conservation of energy which depends on the time duration $\Delta t$ of the interaction $\Delta E \Delta t \gtrsim 1$. In the non-relativistic regime, the uncertainty in the momenta is minimized by choosing $\Delta t \sim \tau_q$, in which case
\begin{equation}
\Delta k_{phys} = \Delta q_{phys} \sim \lambda^{-1/3} m_\phi. \label{lim2}
\end{equation}
Within such uncertainty,} the physical momentum 
\begin{equation}
k_p \equiv k_{phys}(t_c) 
\end{equation}
of the $\phi$ particle at the time of collision can be uniquely reconstructed from the value of the spatial momentum $\tilde p=p$ of the $\chi_2$ particle. 
Indeed, the approximate energy conservation equation
\begin{equation}
\sqrt{m_\phi^2 +k_p^2} + m_\chi \approx \sqrt{m_\psi^2 +(k_p-p)^2} + \sqrt{m_\chi^2+p^2},
\end{equation}
can be written as
\begin{equation}
\sqrt{m_\phi^2 +k_p^2} \approx \sqrt{m_\psi^2 +(k_p-p)^2} + {U\over 2 m_\chi}\label{conse}
\end{equation}
It is then straightforward to show that
\begin{equation}
k_p\approx {M^2 p \over 2U}+ {\rm sign}(p) {\sqrt{M^4 + 4U m_\phi^2} \over 4 m_\chi}, 
\quad\quad {({\rm for}\ k_p \gtrsim \lambda^{-1/3} m_\phi),} \label{k}
\end{equation}
where we have introduced
\begin{equation}
M^2 \equiv m^2_\psi - m^2_\phi + U.
\end{equation}
It can be checked that the double sign in the square root should
be taken as indicated in (\ref{k}), so that Eq.~(\ref{conse}) holds before we square it. 
Thus, we find that there is a single value 
of $k_p$ for each value of $p$, and a measurement of $p$ amounts to a measurement of the momentum of the charged particle at the time of collision. It is also clear from (\ref{k}) that
\begin{equation}
{\rm sign}( k_p) = {\rm sign}(p). \label{lr}
\end{equation}
This is convenient because the sign of $p$ tells us whether our detector has been hit from the left or from the right.

\subsection{Asymmetry}

Let us now look at the distribution (\ref{dist2}), to check for an asymmetry between positive and negative $p$.
First of all, we note that $\psi$ particles are charged and they can also be pair-produced. However, we are only interested in detecting $\phi$ particles, without contamination from $\psi$ particles. Hence, we shall take 
\begin{equation}
\lambda_\psi - \lambda_\phi \gtrsim 1,\label{lplp}
\end{equation}
where we have introduced 
\begin{equation}
\lambda_i = {m_i^2\over eE}. \label{lambdas} 
\end{equation}
Eq.~(\ref{lplp}) ensures that the rate of production of $\psi$ particles is much smaller than the rate of production of $\phi$ particles.
 The amplitude (\ref{amplitu2}) is calculated in Appendix \ref{appB}. Introducing (\ref{liexp}) in (\ref{dist2}), and using
 Eqs. (\ref{ap1}-\ref{ap4}), we have
\begin{equation}
{dN_2(p)\over dt dp} 
= {\hat g^2 \over 16\pi} \left(
\begin{array}{c}
e^{-\pi \lambda_\phi}\\
e^{-\pi \lambda_\psi}\\
\end{array}
\right) {e^{-\pi\sigma_+} \over w_{2} m_\chi U}\ \left|W_{-i\sigma_+,i\sigma_-}\left(i{U\over 2eE}\right)\right|^2,\quad\quad {\rm for}
\left\{\begin{array}{c}
p>0,\\
p<0.\\
\end{array}\right. \label{split}
\end{equation}
Here,
\begin{equation}
\sigma_\pm = {\lambda_\phi \pm \lambda_\psi \over 4}.\label{sigmas}
\end{equation}
In deriving (\ref{split}), we have taken into account that $|{\cal A}_2|$ is independent of $q$. From (\ref{phmo}), this can be understood in terms of the invariance of the interaction probability under time translations. For given $q_{phys}$ the value of $q$ is different depending on the time at which the interaction takes place, so we have  $dN_2/dt dp = (eE) dN_2/dq dp$. 

The most remarkable feature of Eq.~(\ref{split}) is that, together with (\ref{lplp}), it shows a strong asymmetry in the distribution of $p$ between positive and negative momenta:
{
\begin{equation}
{dN_2 (|p|)\over dN_2(-|p|)} = e^{\pi(\lambda_\psi-\lambda_\phi)} \gg 1. \label{asymm}
\end{equation}
If we combine (\ref{asymm}) with (\ref{lr})}, this implies a strong asymmetry in the number of times the collision happens with positive or with negative $k_p$. The smaller is $k_p$, the closer the collision is to the turning point of the $\phi$ particle's hyperbolic trajectory, and the better we test for the frame of nucleation. {In practice, the fundamental uncertainty $\Delta k_p \sim \lambda^{-1/3} m_\phi$ in the momentum of a non-relativistic charged particle will limit the precision to which we can determine the relative velocity between the frame of nucleation and the frame of the detector, so that $\Delta v \sim \lambda^{-1/3}.$
Nonetheless, (\ref{asymm}) is valid beyond the semiclassical regime, and
the distribution of the momentum transfer can  be very strongly biased in the direction of the expanding branch, even if the interaction takes place very close to the turning point.}

{Let us now show that there is a parametric range where the asymmetry (\ref{asymm}) is quite significant, while the collision cannot be described semiclassically.} First, we note that the condition  (\ref{lplp}) for a sizable asymmetry also implies that there is a minimum value of $k_p$, which we may call $k_0$,  for the collision to be kinematically allowed. To find out the value of $k_0$, we can calculate the square of the total 2-momentum 
before and after the collision. The minimum value of $k_p$ is such that the decay products have no relative speed, so
that the total energy is the least possible. Then, in the center of mass frame, the total 2-momentum $P^\mu$ of the products 
of collision only has temporal component, and we have 
\begin{equation}
P^\mu P_\mu \approx (m_\chi + m_\psi)^2, \label{p2}
\end{equation}
where the equality is only approximate because of the fundamental uncertainty in the energy of charged particles.
Before collision, the total 2-momentum in the frame of the $\chi_1$ particle is given by $P^{\mu} \approx \left(m_\chi+\sqrt{m_\phi^2 + k_0^2}, k_0\right)$. Squaring 
and equating to (\ref{p2}) we have
\begin{equation}
k_0^2 \approx (m_\psi^ 2-m_\phi^2) \left[ 1+{m_\psi\over m_\chi}+{m_\psi^2-m_\phi^2 \over 4 m_\chi^ 2}\right].
\end{equation}
It is clear that $k_0$ will be minimized by taking the detector particle to be much heavier than any of the other particles
\begin{equation}
m_\chi \gg m_\psi, m_\phi.
\end{equation}
{
In this case, we have
\begin{equation}
{k^2_{0}} \approx {m_\psi^2-m_\phi^2} \gtrsim \lambda^{-1}_\phi m_\phi^2,\label{momas}
\end{equation}
where the approximate inequality is due to (\ref{lplp}).  From (\ref{k}), this corresponds to momentum transfer of the order 
\begin{equation}
p_0^2 \approx k_0^2 \approx {m_\psi^2-m_\phi^2}.
\end{equation}
On the other hand, the uncertainty in the value of $k_0$ is of order $\Delta k\sim \lambda_\phi^{-1/3} m_\phi.$}

{
We conclude that by choosing $m^2_\psi$ to be in the range
\begin{equation}
\lambda_\phi^{-1} m_\phi^2 \lesssim ({m_\psi^2-m_\phi^2}) \lesssim \lambda_\phi^{-2/3} m_\phi^2,
\end{equation}
and for momentum transfer of the order $p^2\sim p_0^2$, the semiclassical approximation is not valid,
while the asymmetry (\ref{asymm}) in the distribution of $p$ is still quite sizable.
Hence, the asymmetry persists for interactions which occur very close to the turning point of the semiclassical trajectory.}

\section{Summary and conclusions}

We have studied the frame of nucleation of charged particles produced by the Schwinger process,
by using two different model detectors. 

The first model is based on the 3-point interaction (\ref{vertex}) which was introduced in Ref. \cite{GKSSV}. In this case, the detector is a charged particle $\psi$ which anihilates with the $\phi$ antiparticle of a nucleated pair, producing a neutral particle $\chi$.
Semiclassically, the trajectory of a nucleated pair is hyperbolic, with incoming and outgoing branches separated by a turning point. 
In Ref. \cite{GKSSV} it was shown that in the detector's frame, nucleated particles are preferentially encountered after the turning point, with a strong asymmetry between the number of incoming and outgoing particles. The analysis of \cite{GKSSV} concentrated on the detection of particles which are fairly distant from the classical turning point, moving already at ultrarelativistic speed $|v| \approx 1$. In this sense, the relative velocity between the frame of nucleation and the frame of the detector was not very well constrained. Here, we have refined this analysis, extending it to the regime where the $\phi$ particle is non-relativistic at the time of collision. We have shown that the asymmetry persists down to velocities of order $\Delta v\sim S_E^{-1/3}\ll 1$. {Here, $\Delta v$ is the minimum uncertainty in the velocity of a non-relativistic charged particle in the presence of an electric field. This represents a fundamental limitation in the determination of the relative speed, which cannot be overcome.}

The second model we have introduced is based on the 4-point interaction (\ref{vertex4}). Here, the detector is a neutral particle $\chi_1$, which upon interaction with the charged particle $\phi$ produces a second neutral particle $\chi_2$ and the charged particle $\psi$,
{
\begin{equation}
\chi_1 + \phi \to \chi_2 + \psi.
\end{equation}
}
We choose the masses of $\chi_1$ and $\chi_2$ to be the same, so we can think of the neutral sector as a detector of mass $m_\chi$ which is being hit by a charged $\phi$ particle.
In this case, we find a strong asymmetry in the momentum $p$ transferred from the charged to the neutral sector, pointing in the direction of expansion after the turning point [see Eq.~(\ref{asymm})]. Like in the case of the 3-point interaction, this means that throughout the semiclassical regime, the detector is almost exclusively excited by the outgoing (i.e. expanding) branch of the hyperbolic trajectory, and hardly ever by the incoming (i.e. contracting) branch. Moreover, the strong asymmetry persists even for values of $p$ corresponding to interactions which take place near the turning point, where the semiclassical approximation breaks down. We conclude that the rest frame of nucleation is very sharply defined and strongly correlated with the frame of the detector.

There are several directions in which our analysis could be extended. First of all, we have restricted attention to the case where the electric field is switched on for an infinitely long time, and the charged field $\phi$ is in its Lorentz invariant ``in" vacuum. More realistically, the electric field decays by pair production, so its time duration should be finite. In this case, we expect that the frame of nucleation can also be influenced by the frame where the electric field is turned on at some finite initial time. We expect that initial conditions will not play a role provided that the detector's proper time $\tau$ elapsed since the hypersurface of initial conditions is much larger than the time $\tau_{nuc}$ that it takes for a pair to be produced from a vacuum fluctuation.
In Section \ref{spp} we have estimated this time to be of the order of the size of the instanton $r_0$. We therefore expect our results to be valid for $\tau \gg  r_0$.  A more quantitative investigation of this case is left for further research. Also, throughout this paper, we have restricted attention to the case where the detector is a single point-like particle, concluding that the rest frame of nucleation is strongly correlated with the state of motion of the detector. One may then ask what happens if we have an extended detector consisting of, say, two particles in relative motion. In this case we still expect that the frame of nucleation will be strongly correlated with the state of motion of the two particles. The situation is somewhat reminiscent of the Einstein-Podolsky-Rosen (EPR) setup, where a spin zero particle disintegrates into two particles with opposite spin. Once we have measured the spin of one of the products, the measurement of the second one will be maximally correlated with the first. But if the measurements are space-like separated, we cannot say which measurement happened first. Now, when we try to measure the nucleation of a bubble by using two detector particles in relative motion, we also expect that the measurements of the two detectors will be maximally correlated, but it would of course be interesting to find out precisely how this happens.
If the detector particles are described by plane waves, the dominant contribution to scattering amplitudes corresponds to uncorrelated interactions of each one of the particles with pairs which have nucleated far apart from each other. On the other hand, if we consider wave packets localized in position space and restrict the time interval for interaction, then we expect correlations depending on the relative separation of two detector particles. This setup is technically complicated, and we leave it for further research. Another direction which seems worth pursuing is the case of pair production by an electric field in de Sitter space \cite{j}. We expect to come back to this issue in a future publication.

\acknowledgments

{We are grateful to Alex Vilenkin, Jiro Soda and Misao Sasaki for useful comments.}
J.G. and S.K. are grateful to YITP for hospitality during the long term workshop YITP-T-12-03.
This work was supported in part by grant AGAUR 2009-SGR-168, MEC FPA 2010-20807-C02-02, CPAN CSD2007-00042 Consolider-Ingenio 2010, PHY-0855447 from the National Science Foundation, the Grant-in-Aid for Scientific Research (Nos. 21244033, 21111006, 24103006 and 24103001) and the Grant-in-Aid for the Global COE Program “The Next Generation of Physics, Spun from Universality and Emergence” from the Ministry of Education, Culture, Sports, Science and Technology of Japan.

\appendix

\section{Asymptotic expansion of $D_\nu(Z)$}
\label{appA}

The parabolic cylinder functions  of large order $|\nu|\sim \lambda \gg1$ and argument $Z\gtrsim \lambda^{1/2}$ can
be estimated by using an integral representation of $D_\nu(Z)$, and then evaluating this integral by using 
the method of steepest descents.

\begin{figure}
\begin{center}
\vspace{-2cm}
\includegraphics[width=14cm]{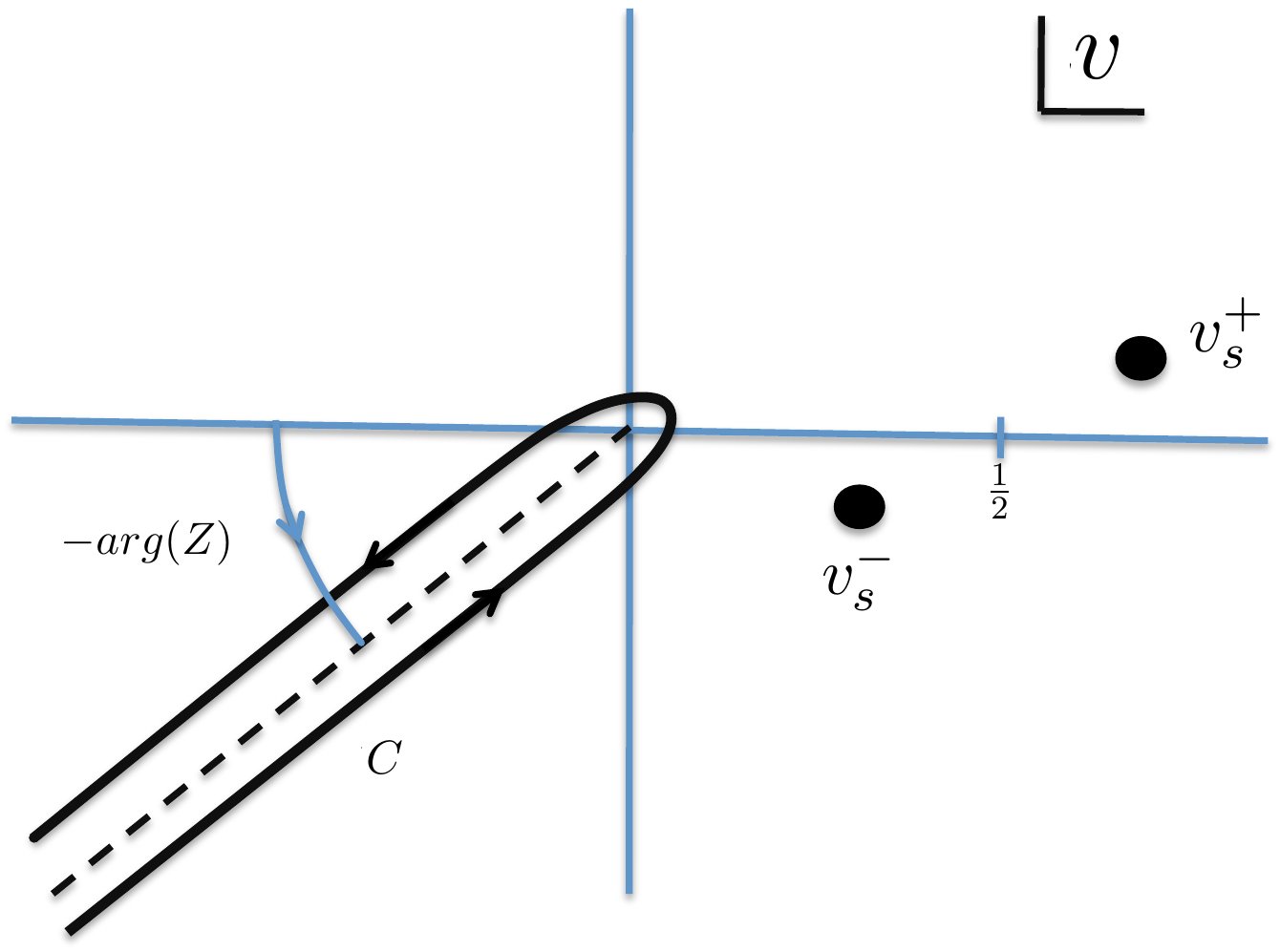}
\vspace{-1.5cm}
\caption{The contour of integration $C$ in Eq.~(\ref{intrep}). The saddle points $v_s^\pm$, given by Eq.~(\ref{vspm})  
are also represented. The root that is closest to $C$ is $v_s^{-}$, and will give the dominant contribution provided that
Eq.~(\ref{sep}) is satisfied. In this case, the saddle point approximation is obtained by deforming $C$ so that it goes
through the point $v_s^-$. \label{fig7}}
\end{center}
\end{figure}

Following \cite{Crothers}, we write 
\begin{equation} 
D_\nu(Z) = {\Gamma(1+\nu) \over 2\pi i} e^{-Z^2/ 4} Z \int_C e^{f(v;Z)} dv, \label{intrep}
\end{equation}
where the contour $C$ is depicted in Fig. \ref{fig7}, and
\begin{equation}
f(v;Z) = Z^2(v-v^2/2)-(1+\nu)\ln(vZ).\label{fex}
\end{equation}
The saddle points are determined by $f'(v_i:Z)=0$, where a prime indicates derivative with respect to $v$. This equation is solved by
\begin{equation}
2 v_s= 1 + {S\over Z}, \label{roots}
\end{equation}
Here we have introduced 
\begin{equation}
S(Z) \equiv \pm [Z^2-4(1+\nu)]^{1/2}.\label{sdef}
\end{equation}
In principle we should consider both possible values of the square root. However,
from the location of the saddle points relative to the overall geometry of the contour, we shall see that the only relevant saddle 
point which is the one closest to the branch cut of $f(v;Z)$ (see Fig. \ref{fig7}). 
From Eqs. (\ref{zdef}) and (\ref{sdef}), we have
\begin{equation}
S^2 = {2\over ieE(4+\alpha^2)} \left[\alpha^2 P^2 - 4(p^2+m_\phi^2) + 4i \alpha p W\right] - 2, \label{s2}
\end{equation}
where
\begin{equation}
P\equiv (W^2-m_\phi^2)^{1/2}.
\end{equation}

By performing the Gaussian integral around the saddle point, we have
\begin{equation}
 \int_C e^{f(v;Z)} dv  \approx \left({2\pi \over -f''_s}\right)^{1/2}\  e^{f_s}\  [1+O(1/f_s'')]. 
 \end{equation}
where
\begin{equation}
f_s'' = - 2 Z^2 {S \over Z+S}
\end{equation}
is the second derivative of $f$ with respect to $v$ at the saddle point, and
\begin{equation}
e^{f_s} = (4e)^{(1+\nu)/2} e^{Z^2/4} e^{ZS/ 4}  (Z+S)^{-(1+\nu)}.
\end{equation}
The saddle point approximation is valid when the distance $|\delta v|=|S/Z|$ between the two saddle points is large enough, so that
the Gaussian integrals around both saddle points do not have much overlap $|f''_s \delta v^2|\gg 1$. This requires
\begin{equation}
|S^3| \gg |S+Z|. \label{sep}
\end{equation}
Substituting in (\ref{intrep}), we have
\begin{equation}
D_\nu(Z) \approx -i \Gamma(1+\nu)\ e^{(1+\nu)/2}  \left({Z+S \over 2}\right)^{-({1\over 2}+\nu)} 
{e^{ZS/ 4} \over \sqrt{2\pi S}},
\end{equation}
from which we find
\begin{equation}
|D_\nu(Z)|^2 \approx {e^{{1\over 2}-\lambda \arg(Z+S)} \over 2 \cosh(\pi\lambda/2)} \left|{e^{ZS/2} \over S}\right|.\label{modsq}
\end{equation}

To proceed, we shall assume  that the difference $m_\chi-m_\psi$ is large enough, so that 
\begin{equation}
\alpha P \gg m_\phi. \label{largeP}
\end{equation}                    
We also assume that $|p|\lesssim P$ {and $\alpha\gg 1$}. From (\ref{s2}), this leads to
\begin{equation}
S\approx \pm \sqrt{2\over ieE} P \left\{1+{2ipW\over \alpha P^2}-{ie E\over 2 P^2}  + O\left[\left({W\over \alpha P}\right)^{2},\left({eE\over P^2}\right)^2\right]\right\}. \label{Sa}
\end{equation}
To lowest order, from (\ref{roots}) we have the two solutions
\begin{equation}
2v_s^{\pm} \approx 1\pm \left({P\over W} +{2i p\over \alpha P} -{ieE\over 2  PW}-{2ipP\over \alpha W^2}\right).\label{vspm}
\end{equation}
It follows that the saddle point which is the closest to the branch cut of the function $f(v;Z)$ given in (\ref{fex}) corresponds to
the lower sign. It is then straightforward to check  that
\begin{equation}
-\lambda \arg(Z+S) + \Re[ZS/2] \approx -4{p P\over \alpha eE} +{\pi\lambda\over 4}-{1\over 2}. \label{expo}
\end{equation}
{Substituting (\ref{modsq}) in (\ref{big}), 
and using (\ref{expo}), we obtain 
\begin{equation}
{dN_\chi }  \approx  {g^2\over 8 eE\ m_\psi}\ {e^{-\pi\lambda}\over   P}\ \exp\left[{- 2\frac{(p+P)^2}{(eET)^2}}\right]{dp\over w_\chi}.
\label{small0}
\end{equation}
Finally, we should summarize the conditions for the validity of (\ref{small0}). Using (\ref{zdef}) and (\ref{Sa}), Eq.~(\ref{sep}) requires 
$P^3 \gg eE\ W$. This is satisfied provided that 
\begin{equation}
P \gtrsim \lambda^{-1/3} m_\phi. \label{minP}
\end{equation}
For smaller values of $P$, the two saddle points are too close together and the Gaussian approximation to the integral representation (\ref{intrep}) is not valid. Also, since we have used (\ref{largeP}) in order to expand $S$ and $v_s^\pm$, we must require
\begin{equation}
P \gg \alpha^{-1} m_\phi,\label{minP2}
\end{equation}
with $\alpha = eET^2 \gg 1$.
}

\section{Transition amplitude for the neutral detector}

\subsection{Probing the ``in'' vacuum state of the $\phi$  field}

\label{appB}

Here we shall calculate the amplitude (\ref{amplitu2}) for the 4-point interaction, starting out in the ``in'' vacuum state for the $\phi$ field. 
For generality, and as a check of Lorentz invariance, here we consider arbitrary momenta $\tilde p'$ and $\tilde p$ for the neutral particles $\chi_1$ and $\chi_2$. 
Then we have
\begin{eqnarray}
{\cal A}_{2} (k,q,{\tilde p}\,;\,{\tilde p}')&=&\frac{\hat g}{(2eE)^{1/2} (2\omega_{\tilde p})^{1/2}(2\omega_{\tilde p'})^{1/2}}~e^{-i\frac{\pi}{4}\nu_\phi-i\frac{\pi}{4}\nu_\psi}
\cr&&\times
\int_{-\infty}^{\infty}dt~e^{i\omega_{\tilde p}t-i\omega_{\tilde p'}t}~
D_{\nu_\phi}[-\sqrt{2} e^{\frac{\pi}{4}i}z_\phi]\,
D_{\nu_\psi}[-\sqrt{2} e^{\frac{\pi}{4}i}z_\psi],
\label{Achi2}
\end{eqnarray}
where we have defined
\begin{eqnarray}
&&z_\phi = \sqrt{eE}\left(t+\frac{k}{eE}\right)\,,
\qquad
\nu_\phi = -\frac{1+i\lambda_\phi}{2}\,,
\qquad
\lambda_\phi = \frac{m_\phi^2}{eE}\,,
\cr
&&z_\psi = \sqrt{eE}\left(t+\frac{q}{eE}\right)\,,
\qquad
\nu_\psi = -\frac{1+i\lambda_\psi}{2}\,,
\qquad
\lambda_\psi = \frac{m_\psi^2}{eE}\,.
\label{variables}
\end{eqnarray}
Using the following integral representation of the parabolic cylinder functions:
\begin{eqnarray}
D_{\nu}[-\sqrt{2} e^{\frac{\pi}{4}i}z]=\frac{e^{-i\frac{\pi}{4}\nu}}{\Gamma(-\nu)}
e^{-\frac{i}{2}z^2}\int_0^{\infty}dx~x^{\nu^*}\,e^{i\sqrt{2}zx-i\frac{x^2}{2}}
\label{cylinder}
\end{eqnarray}
the amplitude becomes
\begin{eqnarray}
{\cal A}_{2} (k,q,{\tilde p}\,;\,{\tilde p}')
&=& 
\frac{\hat g}{(2eE)^{1/2} (2\omega_{\tilde p})^{1/2}(2\omega_{\tilde p'})^{1/2}}~\frac{e^{-i\frac{\pi}{2}\nu_\phi-i\frac{\pi}{2}\nu_\psi}}{\Gamma(-\nu_\phi)\Gamma(-\nu_\psi)}
\nonumber\\
&&\times 
\int_0^\infty dx~ x^{\nu_\phi^*}~ e^{- \, i\frac{x^2}{2}\,+i\sqrt{\frac{2}{eE}}\, kx}
\int_0^\infty dy~ y^{\nu_\psi^*}~ e^{- \, i\frac{y^2}{2}\,+i\sqrt{\frac{2}{eE}}\, qy \, -\frac{i}{2eE}(k^2+q^2)}
\nonumber\\
&&\times  
~\int^\infty_{-\infty} dt~ e^{-ieEt^2-i(\omega_{\tilde p'}-\omega_{\tilde p}+k+q - \sqrt{2eE}\,x-\sqrt{2eE}\,y)\,t}\ .
\end{eqnarray}
After completing the square in the exponent of the $t$-integral and performing the Gaussian integration, we have
\begin{eqnarray}
{\cal A}_{2} (k,q,{\tilde p}\,;\,{\tilde p}')
&=& 
\frac{(2\pi)^{1/2}\hat g}{2eE\, (2\omega_{\tilde p})^{1/2}(2\omega_{\tilde p'})^{1/2}}~\frac{e^{-i\frac{\pi}{2}\nu_\phi-i\frac{\pi}{2}\nu_\psi}}{\Gamma(-\nu_\phi)\Gamma(-\nu_\psi)}~
e^{-\frac{\pi}{4}i}\,e^{-\frac{i}{2eE}(k^2+q^2)+\frac{i}{4eE}(\omega_{\tilde p'}-\omega_{\tilde p}+k+q)^2}
\nonumber\\
&&\times 
\int_0^\infty dx~ x^{\nu_\phi^*} ~e^{-i\mu_+x}
\int_0^\infty dy~ y^{\nu_\psi^*}~ e^{i(x-\mu_-)\,y}
\ ,
\label{Achi2-1}
\end{eqnarray}
where we defined
\begin{eqnarray}
\mu_+=\frac{\omega_{\tilde p'}+{\tilde p'}-(\omega_{\tilde p}+{\tilde p})}{\sqrt{2eE}}
\,,\qquad\quad
\mu_-=\frac{\omega_{\tilde p'}-{\tilde p'}-(\omega_{\tilde p}-{\tilde p})}{\sqrt{2eE}}\,.\label{mupmdef}
\end{eqnarray}
Note that the product of $\mu_+$ and $\mu_-$ is related to the momentum transferred to the
neutral sector at the collision
\begin{equation}
\mu_+\mu_-=-\frac{1}{2eE}\left({\tilde p}-{\tilde p'}\right)^\mu 
\left({\tilde p}-{\tilde p'}\right)_\mu\equiv -{U\over 2eE}.\label{prodmu}
\end{equation}
Defining 
\begin{equation}
u=-\mu_-, \quad \beta= i \mu_+,
\end{equation}
the $y$ integral in Eq.~(\ref{Achi2-1}) is well defined for 
\begin{equation}
0<\arg(u)<\pi, \label{mum}
\end{equation}
and in that case it is given in terms of Euler's Gamma function.
The $x,y$-integrals in Eq.~(\ref{Achi2-1}) then become:
\begin{eqnarray}
\int_0^\infty dx~ x^{\nu_\phi^*} ~e^{-\beta x}
\int_0^\infty dy~ y^{\nu_\psi^*}~ e^{i(x+u)\,y}=
e^{-i\frac{\pi}{2}\nu_\psi}\Gamma(-\nu_\psi)\int_0^\infty dx~
x^{\nu_\phi^*}\left(x+u\right)^{\nu_\psi}~e^{-\beta x}.\label{integral}
\end{eqnarray}
Likewise, the existence of the integral in the right hand side of Eq.~(\ref{integral}) requires
\begin{equation}
-{\pi\over 2} <\arg(\beta) < {\pi\over 2}.\label{mup}
\end{equation} 
From Eq. 3.383.4 in Ref. \cite{GR} we have
\begin{equation} 
\int_0^\infty dx~
x^{\nu_\phi^*}\left(x+u\right)^{\nu_\psi}~e^{-\beta x}= \beta^{-{1\over 2}-i\sigma_-} u^{-{1\over 2}+i\sigma_-} \Gamma(-\nu_\phi) 
\ e^{\beta u\over 2}\ W_{-i\sigma_+,i\sigma_-} (\beta u). \label{GR}
\end{equation}
Here, we have introduced 
\begin{equation}
\sigma_{\pm} \equiv \frac{\lambda_\phi\pm\lambda_\psi}{4}\,
\end{equation}
Eq.~(\ref{GR}) is valid for real positive values of the parameters $\beta >0$ and $u>0$,
but it can be analytically continued to the domain 
\begin{equation}
\Re \beta >0, \quad |\arg(u)| < \pi. \label{domain}
\end{equation}
If $u$ were real and negative, the contour of integration would step over a singularity at $x=-u$,
so the integral has a branch cut at $u<0$. On the other hand, the domain (\ref{domain}) includes the parameter region of our interest, which is given by (\ref{mum}) and (\ref{mup}). 

Using Eqs. (\ref{Achi2-1}), (\ref{integral}) and (\ref{GR}), we have 
\begin{equation}
|{\cal A}_{2} ({\tilde p}\,;\,{\tilde p}')|^2
=\hat g^2
\frac{\pi e^{-\frac{\pi}{2}\lambda_\phi-\pi\lambda_\psi}}
{8(eE)^2\, \omega_{\tilde p}\omega_{\tilde p'}}{|\beta^{-i\sigma_-}u^{+i\sigma_-}|^2 \over 
|\mu_-\mu_+|} {|W_{-i\sigma_+,i\sigma_-} (\beta u)|^2 }, \label{liexp}
\end{equation}
Now, using (\ref{mup}) and (\ref{mum}) we can disambiguate the phases of $\beta$ and $u$ as follows.
\begin{equation}
\beta = |\mu_+| \left(
\begin{array}{c}
e^{i {\pi\over 2}}\\
e^{-i{\pi \over 2}}\\
\end{array}
\right),\quad {\rm for}\left\{
\begin{array}{c}
\mu_+>0,\\
\mu_+<0.\\
\end{array}
\right.\label{ap1}
\end{equation}
and
\begin{equation}
u = |\mu_-| \left(
\begin{array}{c}
e^{i {\pi}}\\
1\\
\end{array}
\right),\quad {\rm for}\left\{
\begin{array}{c}
\mu_->0,\\
\mu_-<0.\\
\end{array}
\right.\label{ap2}
\end{equation}
From (\ref{mupmdef}) and (\ref{prodmu}), it is clear that $\mu_+$ and $\mu_-$ are the null components
of the momentum transfer 2-vector, and therefore the signs ${\rm sign}(\mu_\pm)$ and the product 
$(\mu_+\mu_-)$ are Lorentz invariant. Hence, Eq.~(\ref{liexp}) for the modulus squared of the amplitude is written in a manifestly Lorentz invariant form. 

In the frame where the initial particle is at rest, we have $\tilde p'=0$, $\tilde p=p$. Moreover, in  
the case when the masses of the neutral partilces are degenerate, $m_1=m_2=m_\chi$, we have
\begin{equation}
{\rm sign}(\mu_+)=-{\rm sign}(\mu_-)=-{\rm sign}(p), \label{ap3}
\end{equation}
while from (\ref{prodmu}) we have 
\begin{equation}
|\mu_+\mu_-| = {U \over 2eE}>0. \label{ap4}
\end{equation}
Introducing (\ref{liexp}) in (\ref{dist2}), and using
 Eqs. (\ref{ap1}-\ref{ap4}), we derive Eq.~(\ref{split}) in the main text.

 \subsection{Probing a one particle state for the $\phi$ field}

\label{appB2}

{As a consistency check, and an alternative way of deriving the momentum discontinuity in Eq.~(\ref{split}),} let us now ``calibrate" our 4-point interaction detector by introducing an actual $\phi$ particle of momentum $k$ in the initial state, in addition to the neutral $\chi_1$ particle. 
Changing $\phi_{k}(z)$ to $\phi^*_{k}(z)$ in Eq.~(\ref{amplitu2}), the new amplitude is given by
\begin{eqnarray}
\hat{\cal A}_{2} (k,q,{\tilde p}\,;\,{\tilde p}')&=&\frac{\hat g}{(2eE)^{1/2} (2\omega_{\tilde p})^{1/2}(2\omega_{\tilde p'})^{1/2}} 
{e^{i\frac{\pi}{2} (\nu^*_\phi-\nu_\psi)} \over \Gamma(-\nu^*)\Gamma(-\nu_\psi)}\\
&\times& 
\int_0^\infty dx~ x^{\nu_\phi}~ e^{ \, i\frac{x^2}{2}\,-i\sqrt{\frac{2}{eE}}\, kx}
\int_0^\infty dy~ y^{\nu_\psi^*}~ e^{- \, i\frac{y^2}{2}\,+i\sqrt{\frac{2}{eE}}\, qy \, +\frac{i}{2eE}(k^2-q^2)}
\nonumber\\
&&\times  
~\int^\infty_{-\infty} dt~ e^{-i(\omega_{\tilde p'}-\omega_{\tilde p}-k+q +\sqrt{2eE}\,x-\sqrt{2eE}\,y)\,t}\ .
\end{eqnarray}
Using $q+\tilde p= k+ \tilde p'=0$, the delta function which results from time integration can be written as 
\begin{equation} 
\int^\infty_{-\infty} dt~ e^{-i(\omega_{\tilde p'}-\omega_{\tilde p}-k+q +\sqrt{2eE}\,x-\sqrt{2eE}\,y)\,t} = {2\pi \over (2eE)^{1/2}} \delta(\mu_+ +x-y).
\end{equation}
For $\mu_+>0$ we do the $y$ integral, and after using Eq.~(\ref{GR}), we obtain:
\begin{equation}
|\hat{\cal A}_{2} |^2= {\pi^2 \hat g^2 \over 4 (eE)^2 w_{\tilde p} w_{\tilde p'}}{e^{-2\pi\sigma_+ +\pi \sigma_-} \over |\Gamma(-\nu_\psi)|^2} {|W_{+i\sigma_+,-i\sigma_-}(i\mu_+\mu_-)|^2 \over |\mu_+\mu_-|},  
\quad\quad (\mu_+>0), \label{cali1}
\end{equation} 
while for $\mu_+<0$, we must do the $x$ integral first, and we obtain
\begin{equation}
|\hat{\cal A}_{2} |^2= {\pi^2 \hat g^2 \over 4 (eE)^2 w_{\tilde p} w_{\tilde p'}}{e^{-2\pi\sigma_+ - \pi \sigma_-} \over |\Gamma(-\nu^*_\phi)|^2} {|W_{-i\sigma_+,-i\sigma_-}(-i\mu_+\mu_-)|^2 \over |\mu_+\mu_-|},
\quad\quad (\mu_+<0). \label{cali2}
\end{equation} 
Note, however, that $W_{\mu,\nu}(z)=W_{\mu,-\nu}(z)$, and so the factor containing the Whittaker function is common in Eqs. (\ref{cali1}) and (\ref{cali2}). Finally, setting $\tilde p =p$ and $\tilde p'=0$, we have 
\begin{equation}
 |\hat{\cal A}_{2} |^2 = {\pi \hat g^2 \over 4 eE} \left[
1+ 
 \left(
\begin{array}{c}
e^{-\pi \lambda_\phi}\\
e^{-\pi \lambda_\psi}\\
\end{array}
\right) \right] {e^{-\pi \sigma_+} \over m_\chi w_2 U}\ \ 
\left|W_{-i\sigma_+,i\sigma_-}\left(i{U\over 2eE}\right)\right|^2,\quad\quad {\rm for}
\left\{\begin{array}{c}
p>0,\\
p<0.\\
\end{array}\right. \label{split2}
\end{equation}
{The dominant term in (\ref{split2}) is symmetric in $p$, and corresponds to the fact that the initial $\phi$ particle is in hyperbolic motion. For any given $|p|$, its trajectory has two chances for interacting with the detector. In one case the detector is hit from the right, and in the other case from the left, which gives rise to the two possible signs of $p$ with equal probability. The subleading term in (\ref{split2}), on the other hand, has the same momentum asymmetry as the distribution (\ref{split}). This is not surprising, since aside from the original $\phi$ particle of momentum $k$, we will have a bath of pairs of $\phi$ and $\psi$ particles which are created by the electric field. In particular, there will be a pair of $\phi$ particles of momentum $k$ which will be created, with probability proportional to $e^{-\pi\lambda_\phi}$. Upon interaction, the created $\phi$ particle will deposit positive momentum transfer $p>0$ on the detector. Likewise, a pair of $\psi$ particles of momentum $q$ can be created with probability proportional to $e^{-\pi\lambda_\psi}$, and upon interaction, the $\psi$ antiparticle will deposit negative momentum transfer $p<0$ on the detector.}

\end{document}